\documentclass[aps,prd,twocolumn,showpacs,eqsecnum,nofootinbib,floatfix]{revtex4-1}
\pdfoutput=1
\usepackage{amssymb}
\usepackage{graphicx}
\usepackage{amsmath}

\font\FermiSmallfont=cmssq8 scaled 1200

\def\UMDppthead#1#2#3{
\null 
\begin{center}\vskip -1.0truein{\hbox to 7.5truein {
\hfill
\vbox to 1in {\vfill \FermiSmallfont
              \hbox{#1}
              \hbox{#2}
              \hbox{#3}
              \vfill}
}}\vskip-0.0truein\end{center}}%FNALppthead

\def\apj{Astrophys.\ J.}

\begin{document}

%\preprint{UMD-PP-11-012}

\UMDppthead{UMD-PP-11-012}{}{}

\title{The Halo Model of Large Scale Structure for Warm Dark Matter}

\author{Robyn M.\ Dunstan$^1$}\email{rdunstan@umd.edu}
\author{Kevork N.\ Abazajian$^{1,2}$}\email{kevork@uci.edu}
\author{Emil Polisensky$^3$}\email{Emil.Polisensky@nrl.navy.mil}
\author{Massimo Ricotti$^4$}\email{ricotti@astro.umd.edu}
\affiliation{$^1$Maryland Center for Fundamental Physics \& Joint
  Space-Science Institute, Department of Physics, University of
  Maryland, College Park, Maryland 20742 USA}
\affiliation{$^2$Department of Physics \& Astronomy, University of California, Irvine, CA 92697 USA}
\affiliation{$^3$Naval Research Laboratory, Washington, D.C. 20375, USA} 
\affiliation{$^4$Department of Astronomy, University of
  Maryland, College Park, Maryland 20742 USA}

\date{\today}

\begin{abstract}
  We present a comprehensive analysis of the halo model of
  cosmological large to small-scale structure statistics in the case
  of warm dark matter (WDM) structure formation scenarios.  We include
  the effects of WDM on the linear matter power spectrum, halo density
  profile, halo concentration relation, halo mass function, subhalo
  density profile, subhalo mass function and biasing of the smooth
  dark matter component. As expected, we find large differences at the
  smallest physical scales in the nonlinear matter power spectrum
  predicted in the halo model between WDM and cold dark matter even
  for reasonably high-scale WDM particle masses.  We find that
  significant effects are contributed from the alteration of the halo
  density profile and concentration, as well as the halo mass
  function.  We further find that the effects of WDM on the subhalo
  population are important but sub-dominant.  Clustering effects of the
  biasing of the smooth component in WDM is not largely significant.
\end{abstract}


%\keywords{}

\maketitle

\section{Introduction}
\label{intro}

The theory of large scale structure growth and formation from
Gaussian, adiabatic initial conditions is in remarkable agreement with
observed large scale structure in the universe at scales above the
Galactic scale.  With the success of the Wilkinson Microwave
Anisotropy Probe ~\cite{Komatsu:2010fb} at determining the amplitude
and shape of the matter power spectrum, and upcoming results from the
Planck Probe, the primordial linear clustering of matter will be known
to better than 1.1\% in the amplitude and 0.22\% in the slope of the
power spectrum.  When combined with galaxy surveys such as the Sloan
Digital Sky Survey (SDSS) and 2-degree Field, the precision can be
even further enhanced \cite{Burigana:2010hg}.  These precise
determinations of the clustering power spectrum, coupled with the
ansatz of {\em cold} dark matter (CDM) with a primordial power law
spectrum leads to a precise prediction for the power spectrum to
arbitrary small scales of relevance to galaxy formation.

However, the small-scale clustering of dark matter can be largely
different at the small scales inaccessible to CMB measurements as well
as galaxy surveys, yet are significant in the role of galaxy
formation.  The small-scale clustering of matter can be suppressed by
thermal motion of the dark matter, i.e., {\em warm} dark matter (WDM),
allowing the dark matter to free stream out of primordial potential
wells.  It can also be altered in extended inflationary models
~\cite{Kamionkowski:1999vp}, and in the case of charged or neutral
decaying~\cite{Sigurdson:2003vy,*Kaplinghat:2005sy,*Cembranos:2005us}
dark matter progenitor particles.

The suppression of power on small-scales has been proposed as a
solution to a number of problems in galaxy formation: first, the
reduction of satellite galaxy halos
\cite{Kauffmann:1993gv,*Klypin:1999uc,*Moore:1999wf,*Willman:2004xc},
second, the reduction of galaxies in
voids~\cite{Peebles:2001nv,Bode:2000gq}, third, the low concentrations
of dark matter in
galaxies~\cite{Dalcanton:2000hn,*vandenBosch:2000rz,*Zentner:2002xt,*Parry:2011iz},
fourth, the angular momentum problem of galaxy
formation~\cite{Dolgov:2001nq}, fifth, the formation of disk-dominated
galaxies \cite{Governato:2002cv,*Kormendy2005}.

The discovery of a number of dwarf galaxies with the SDSS, and their
confirmation as harboring massive dark matter halos
\cite{Simon:2007dq}, alleviates the first problem, and even more,
constrains the excessive suppression present in WDM models
\cite{Polisensky:2010rw}.  In addition, the small-scale power spectrum
of dark matter inferred from the Lyman-$\alpha$ forest toward distant
quasars is consistent with CDM down to sub-dwarf galaxy
scales~\cite{Seljak:2006qw,*Viel:2007mv}, though the thermal history
of the intergalactic medium consistent with these findings is
difficult to reconcile with other
measures~\cite{Ricotti:1999hx,*Schaye:1999vr}.  Overall, however,
there is considerable interest in WDM as a potential structure
formation scenario, and as a method to test the CDM ansatz.

In this paper, we construct a complete analytic model for the
clustering of dark matter in two-point statistical measures based on
the halo model of large scale structure. For a review of the halo
model, see Ref.~\cite{Cooray:2002dia}.  The forms developed here may
be extended to higher order statistics as well.  The halo model has
been shown to be an effective and accurate model of dark matter
clustering built upon analytic and numerical structure formation
statistics. And, it can be extended to models of
galaxy~\cite{Berlind:2001xk}, gas~\cite{Cooray:2000xh} and galaxy
velocity statistics~\cite{Sheth:1995is}.  It has been shown to be
consistent with broad classes of galaxy clustering measures over a
wide range of redshifts, c.f. Ref.~\cite{Cooray:2005mm}.

The halo model has been employed to investigate the effect of
baryons~\cite{Zhan:2004wq,*Rudd:2007zx} and massive active
neutrinos~\cite{Abazajian:2004zh,*Hannestad:2006as} on weak lensing
statistics.  Recently, the effects of WDM in weak lensing has been
studied through estimates of WDM's alterations of limited components
of the halo model. In Ref.~\cite{Markovic:2010te}, the effects of WDM
suppression of the linear power spectrum and a suppression of the
small-mass slope of the halo mass function were incorporated into an
estimate of the sensitivity of future weak lensing surveys to WDM. In
Ref.~\cite{Smith:2011ev}, the WDM effects of the lack of small-mass
halos in the mass function, potential cores in the halo density
profile, and biasing of the smooth component were also included in
estimates of the sensitivity to WDM in weak lensing surveys.

In this paper, we incorporate results from our numerical simulations
of WDM structure formation as well as the leading results from CDM
halo statistics to construct a complete halo model of large scale
structure in the case of WDM.  We include new numerical simulations'
measures of the halo mass function, halo profile, and sub-halo mass
function in this work.  Importantly, since WDM is known to strongly
affect halo substructure, we include the effects of halo substructure
suppression here, which has not been included in prior work.  We also
incorporate the change of the halo-profile density slope and the
halo-profile concentration relation, which are known to depend on the
initial matter power spectrum~\cite{Bullock:1999he,Ricotti:2002qu},
as well an estimate of the biasing of the smooth dark matter component in
WDM similar to that in Ref.~\cite{Smith:2011ev}.  Notably, we do not
include cored halos in the case of WDM since these have not been
observed in our simulations or others, nor are significant cores
expected from analytic Gaussian-peak statistics
methods~\cite{deNaray:2009xj,VillaescusaNavarro:2010qy}.

Two primary candidates for WDM particles are
gravitinos~\cite{Pagels:1981ke} and sterile
neutrinos~\cite{Dodelson:1993je}.  For concreteness, we employ the
popular Dodelson-Widrow (DW)~\cite{Dodelson:1993je} thermally-produced
sterile neutrino dark matter as our WDM particle dark matter
candidate.  All particle dark matter masses to which we refer are that
of the DW sterile neutrino, unless stated otherwise.

%include WDM problems at explaining core issues, Manoj \& James, 

%What we set cosmological parameters to

%Main equation, then say "and these are the things we changed"

\section{Simulations}
\label{sims}

We employ three sets of simulations using cosmological parameters
consistent with the third year WMAP release~\cite{Spergel:2006hy},
spectral index $n = 0.951$, matter density $\Omega_m = 0.238$, baryon
density $\Omega_b = 0.04$, neutrino density $\Omega_{\nu} = 0$, Hubble
parameter $h = 0.73$, and mass fluctuation with $R = 8 h^{-1}$~Mpc,
$\sigma_8 = 0.751$. We assume all matter in the simulations is dark
matter only but we use $\Omega_b$ for calculating the effects of
baryons on the matter power spectrum. The simulations are conducted
with the $N$-body code GADGET-2~\cite{Springel:2005mi} with initial
conditions generated with the GRAFIC2 software
package~\cite{Bertschinger:2001ng}. Our initial conditions include
particle velocities due to the gravitational potential using the
Zeldovich approximation but we do not add random thermal velocities
appropriate for WDM to the simulation particles. Gravitationally bound
dark matter halos are identified using the AMIGA's Halo Finder
software \cite{Knollmann:2009pb}. Each simulation set consists of a
single realization of the density field but with varied power spectra
of fluctuations appropriate for cold dark matter (CDM) and WDM
cosmologies. We use the WDM transfer function in
Ref.~\cite{Bode:2000gq} valid for gravitino particles. In
Ref.~\cite{Viel:2005qj} the transfer function of DW sterile neutrinos
is shown to be nearly identical to that for gravitinos but with a
scaling relationship between the particles masses. We characterize our
WDM simulations in terms of the DW sterile neutrino mass.

To investigate the halo mass function our first simulation set
consists of $512^3$ particles in a $60^3$~Mpc$^3$ comoving box with
mass resolution $5.7\times10^7 M_{\odot}$ and force resolution
$1.2$~kpc. We run simulations for CDM and WDM cosmologies with sterile
neutrino particle masses of 0.2, 0.7, 1.7, and 4.4~keV.

We use the high resolution simulations described in
Ref.~\cite{Polisensky:2010rw} to investigate the subhalo mass
function. These simulations use a `zoom' technique to sample small
volumes with fine mass resolution containing two Milky Way-sized
halos, M~$\sim 2\times10^{12} M_{\odot}$, embedded in coarsely sampled
larger volumes. The coarse volume is a $90^3$~Mpc$^3$ comoving box
with $\sim 510^3$ particles with mass and force resolutions
$9.2\times10^4 M_{\odot}$ and 275~pc in the refinement region. These
simulations include CDM and 4.4, 11, and 28~keV sterile-neutrino particle mass
WDM cosmologies.

The purpose of our final simulation set is to investigate the halo
density profile. No differences were observed in the density profile
slopes of the Milky Way-sized halos simulations, however these halos
are well above the scale of power suppression in the WDM
cosmologies. To reduce the halo mass to $\sim 10^8 M_{\odot}$ we
selected one of these halos for resimulation in a smaller box,
$4.5^3$~Mpc$^3$, with $\sim 255^3$ particles with mass and force
resolutions of $92 M_{\odot}$ and 14~pc in the refinement region. We
perform simulations for CDM and 28, 48, and 70~keV standard
sterile-neutrino particle WDM cosmologies. We examine the halo density
profiles at redshift 1.08 where the halo appears to be well relaxed in
all cosmologies.

\section{\label{linPk}Matter Power Spectrum}

\subsection{\label{linPk:CDM}Cold Dark Matter}

%Do how one gets the power spectrum from CDM model
The cold dark matter (CDM) power spectrum has been well-defined in the
halo model of large scale structure (e.g.,
Ref.~\cite{Cooray:2002dia}).  The dimensionless power spectrum is
\begin{equation}
  \Delta^2\left(k\right)\equiv\frac{k^3P\left(k\right)}{2\pi^2}, 
\end{equation}
and the mass variance within region $R$ is $\sigma^2\left(R\right)$,
\begin{equation}
\label{eqn:sig} 
\sigma^2\left(R\right) = \int
  \frac{dk}{k} \Delta^2\left(k\right)\hat{W}^2\left(kR\right), 
\end{equation}
where $\hat{W}\left(kR\right)$ is the Fourier transform of the top-hat
function with radius $R$.  The quantity $\sigma$ is also given in
terms of the halo mass given the mass enclosed in radius $R$ for the
mean density of the universe.

We take the linear matter power spectrum for cold dark matter from
Ref.~\cite{Eisenstein:1997jh}.  The power spectrum is the primordial
power spectrum modified by the appropriate transfer function,
\begin{equation} 
\Delta^2\left(k,z\right)=\delta^2_\text{H}\left(\frac{ck}{H_0}\right)^{3+n}T^2\left(k,z\right),
\end{equation}
where $T\left(k,z\right)$ is the CDM transfer function, $\delta_H$ is
the amplitude of perturbations on the scale of today's horizon (and is
absorbed into the normalization constant), and $n$ is the initial
power spectrum index.  We employ the transfer function for CDM plus
baryons from Ref.~\cite{Eisenstein:1997jh}.  This transfer function is
sufficiently accurate for our purposes.  The cosmological parameters
we use for the transfer functions and analytic halo model are:
spectral index $n=1$, CMB temperature $T_\text{CMB}=2.726$, dark
matter density $\Omega_\text{dm}=0.23$, baryon density
$\Omega_\text{b}=0.04$, neutrino density $\Omega_\nu=0$, Hubble
parameter $h=0.7$, and mass fluctuation with $R=8\ h^{-1}\text{ Mpc, }
\sigma_8=0.8$, which are consistent with WMAP7~\cite{Komatsu:2010fb}.
We explore results at redshift $z=0$.

\subsection{\label{linPk:WDM}Warm Dark Matter}

\begin{figure}
 \includegraphics[width=\linewidth]{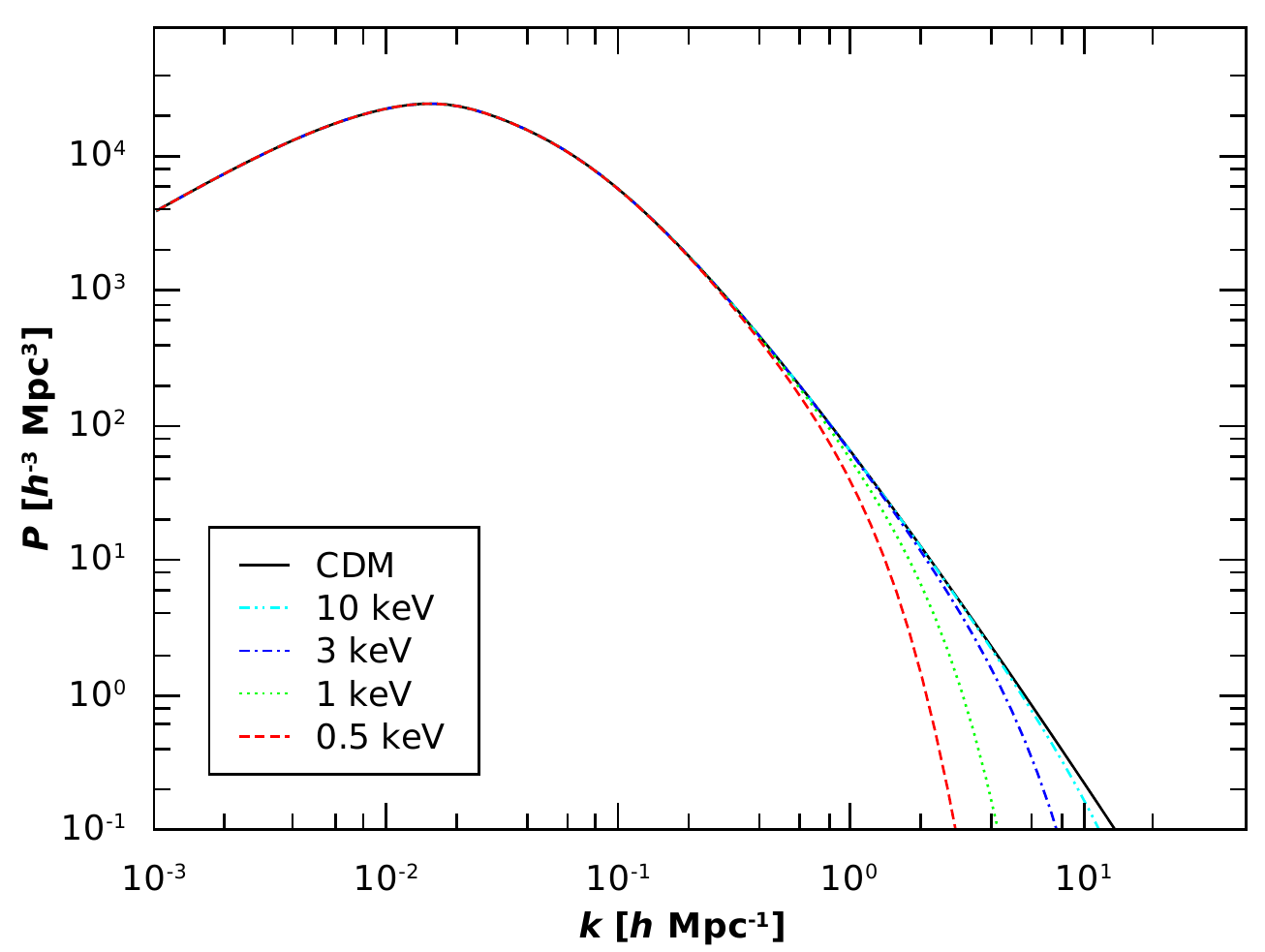}%
 \caption{\label{img:Plin}Shown is the linear matter power spectrum as
   a function of wavenumber for the CDM model and several WDM particle
   masses, as listed in the legend. The smaller the WDM particle mass,
   the greater the suppression of the power spectrum on small scales
   (large $k$).}
 \end{figure}

For the WDM case, for concreteness, we use the sterile neutrino
transfer function from Ref.~\cite{Abazajian:2005xn}.  There is a
simple scaling relation to connect between sterile neutrinos and
gravitino WDM transfer functions given in
Ref.~\cite{Abazajian:2005xn}.  The sterile neutrinos have an initial
velocity dispersion, which allows them to escape the gravitational
well of the small scale perturbations upon entering the horizon.  This
suppresses the linear matter power spectrum for small scales (large
$k$), as shown in Fig.~\ref{img:Plin}. The scale and magnitude of
the suppression increases with velocity and, therefore, decreases with
the mass of the sterile neutrino.  The linear matter power spectrum
for the suppressed sterile neutrino case at small scales is related to
that in CDM by:
\begin{equation} 
  T_\text{s}\left(k\right) \equiv
  \sqrt{\frac{P_\text{sterile}\left(k\right)}{P_\text{CDM}\left(k\right)}}. 
\end{equation}

 In Ref.~\cite{Abazajian:2005xn}, the linear matter power spectrum for
 sterile neutrinos was calculated using CAMB
 \footnote{http://camb.info} using a nonthermal momentum distribution
 and the evolution equations for massive neutrinos.  The resulting fit
 to the sterile neutrino transfer function is
\begin{equation}
\label{eqn:T_wdm} 
T_\text{s}\left(k\right) = \left(1+\left(\alpha
    k\right)^\nu\right)^{-\mu}, 
\end{equation}
where $\nu=2.52$, $\mu=3.08$, and $\alpha$ is a function of the
sterile neutrino mass,
\begin{equation}\label{eqn:Ts_alpha} 
  \alpha = a \left(\frac{m_\text{s}}{1\:\text{keV}}\right)^b
  \left(\frac{\Omega_\text{dm}}{0.26}\right)^c 
  \left(\frac{h}{0.7}\right)^d h^{-1}\:\text{Mpc},
\end{equation}
where $a=0.188$, $b=-0.858$, $c=-0.136$, and $d=0.692$.

\subsection{\label{linPK:Pnl}Nonlinear Matter Power Spectrum}

The nonlinear matter power spectrum consists of two parts: the one
halo term and the two halo term~\cite{Cooray:2002dia}.  These are
denoted as $P_\text{1h}$ and $P_\text{2h}$, respectively:
\begin{equation} 
  P\left(k\right) =
  P_\text{1h}\left(k\right)+P_\text{2h}\left(k\right).
\end{equation}
The one halo term is Fourier transform of the two-point correlation
function for two points that are inside the same halo:
\begin{equation}
\label{eqn:1h} 
P^\text{1h}\left(k\right)=\int dM
\frac{dn}{dM}\left(\frac{M}{\bar{\rho}}\right)^2\left|u\left(k\mid
    M\right)\right|^2. 
\end{equation}
The one halo term depends only on the halo mass function and the halo
density profile.  The two halo term is the Fourier transform of the
two-point correlation function for points that are in different halos:
\begin{eqnarray}\label{eqn:2h}
  P^\text{2h}\left(k\right)&=&\int dM_1\frac{dn}{dM_1} \frac{M_1}{\bar{\rho}}u\left(k\mid M_1\right) \\
  &&\times\int dM_2\frac{dn}{dM_2} \frac{M_2}{\bar{\rho}}u \left(k\mid M_2\right) \nonumber \\
  &&\times P_\text{hh}\left(k\mid M_1,M_2\right). \nonumber
\end{eqnarray}
In addition to dependence on the mass and density functions, the two
halo term also depends on the power spectrum of halos with masses
$M_1\text{ and }M_2\ \left(P_\text{hh}\right).$ $P_\text{hh}$ can
be approximated by the bias for each halo and the linear matter power
spectrum:
\begin{equation} P_\text{hh}\left(k\mid M_1,M_2\right)\approx
  b_1\left(M_1\right)b_2\left(M_2\right)P_\text{lin}\left(k\right). 
\end{equation}

\begin{figure}
\includegraphics[width=\linewidth]{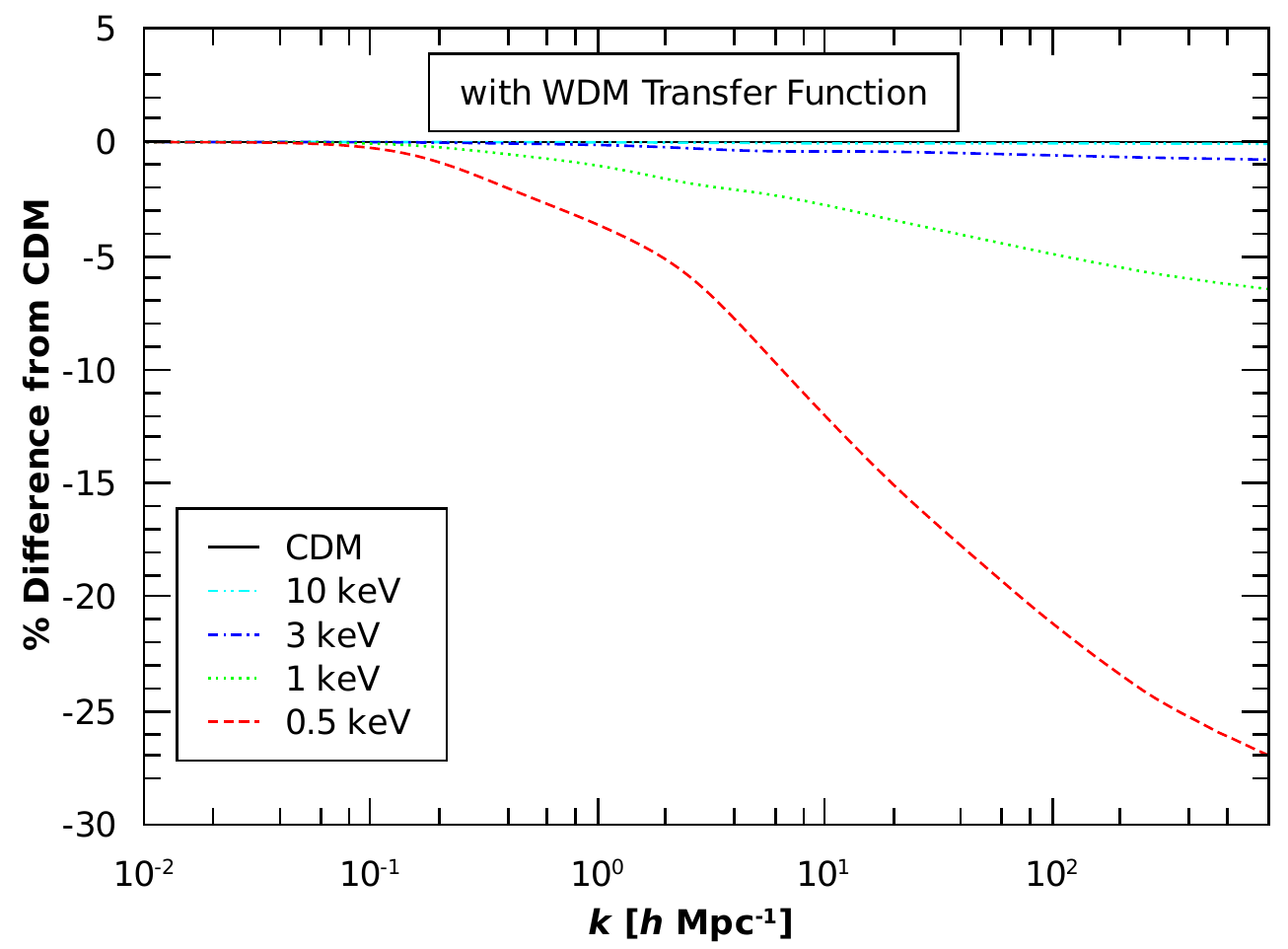}
\caption{\label{img:Pnl_2}Shown is the difference between the WDM
  models and CDM for the nonlinear matter power spectrum with only a
  change in the linear matter transfer function. The WDM particle
  masses are listed in the legend.  This results in suppression of the
  power spectrum on small scales (large $k$) in the WDM models.}
\end{figure}

In the nonlinear matter power spectrum, the suppression of the linear
matter power spectrum only directly effects the 2-halo term.  As with
the linear matter power spectrum, the nonlinear spectrum is suppressed
at small scales, as shown in Fig.~\ref{img:Pnl_2}.

\section{\label{haloMF:bias}Halo Bias Function}

The halo bias function describes how halos cluster relative to the
matter power spectrum.  It is defined as the ratio of the halo power
spectrum to the linear power spectrum~\cite{Tinker:2010my},
\begin{equation} 
b^2\left(k\right) =
  \frac{P_\text{hh}\left(k\right)}{P_\text{lin}\left(k\right)}. 
\end{equation}

\begin{figure}
\includegraphics[width=\linewidth]{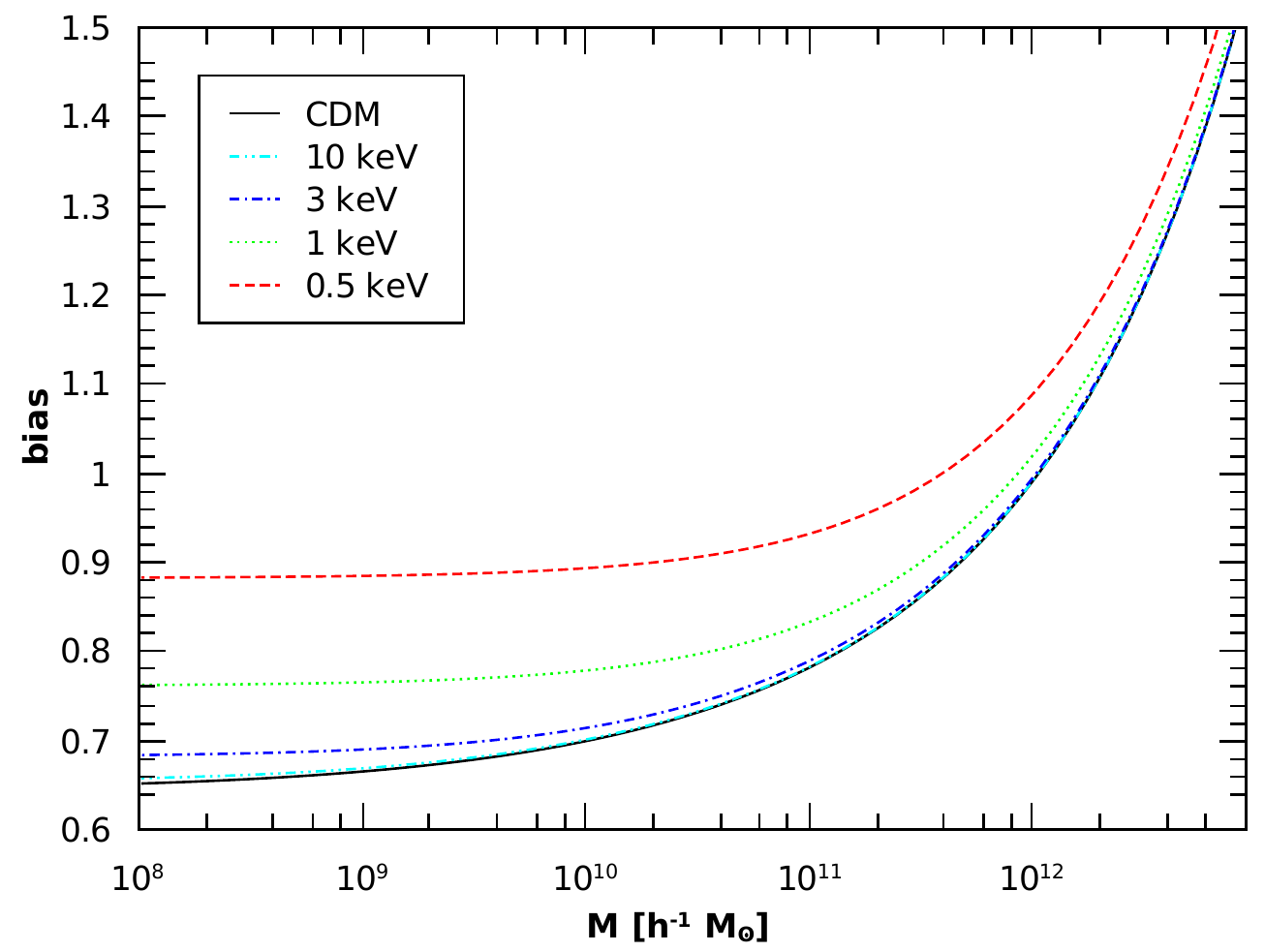}
\caption{
  \label{img:bias}Shown is the halo bias as a function of mass for CDM
  and several WDM particle masses, as given in the legend.  The
  difference in the WDM models comes only from the change in
  $\sigma\left(M\right)$.}
\end{figure}

As shown in Fig.~\ref{img:bias}, small mass halos are less strongly
biased than higher mass halos.  Ref.~\cite{Sheth:1999mn} showed that
halos with $M<M_\star$ are more strongly clustered and halos with
$M>M_\star$ are less strongly clustered than given from the bias
function based on the standard Press-Schechter
formalism~\cite{Press:1973iz}.  $M_\star$ is the typical mass scale
of halos that are currently collapsing.  Ref.~\cite{Seljak:2004ni}
also studied the halo bias function.  They find that previous forms of
the bias function overestimate the bias for halo masses slightly less
than $M_\star$ and that the bias is approximately constant for small
mass halos $\left(M < 0.1 M_\star\right)$.

The halo bias function using a larger suite of simulations consistent
with extended Press-Schechter formalism was found in
Ref.~\cite{Tinker:2010my}, which we employ here.  Specifically, we use
the bias function in \S 3.1, Eq.~(6):
\begin{equation} 
b\left(\nu\right) =
  1-A\frac{\nu^a}{\nu^a+\delta_\text{c}^a}+B\nu^b+C\nu^c 
\end{equation}
with parameters given in Ref.~\cite{Tinker:2010my}, Table 2, where
$\nu$ is defined by
\begin{equation}
\label{eqn:nu} 
\nu =
  \frac{\delta_\text{c}}{\sigma\left(M\right)} 
\end{equation}
where $\sigma\left(M\right)$ is the variance at a given mass scale.

The quantities $\delta_\text{c}$ and $M_\star$ are defined in the
spherical collapse model (see, e.g. Ref.~\cite{Cooray:2002dia}).
Here, $M_\star\left(z\right)$ is defined as the typical mass that is
collapsing at redshift $z$.  The value for $M_\star$ is given by
\begin{equation}\label{eqn:Mstar}
  \sigma\left(M_\star\left(z\right)\right) =
  \frac{\delta_\text{c}}{D\left(z\right)}, 
\end{equation}
where $D$ is the linear growth, and note that $D\left(0\right)=1$.
The constant $\delta_\text{c}$ is the density for collapse in the
spherical collapse model, and is $\delta_\text{c}=1.69$.

\begin{figure}
\includegraphics[width=\linewidth]{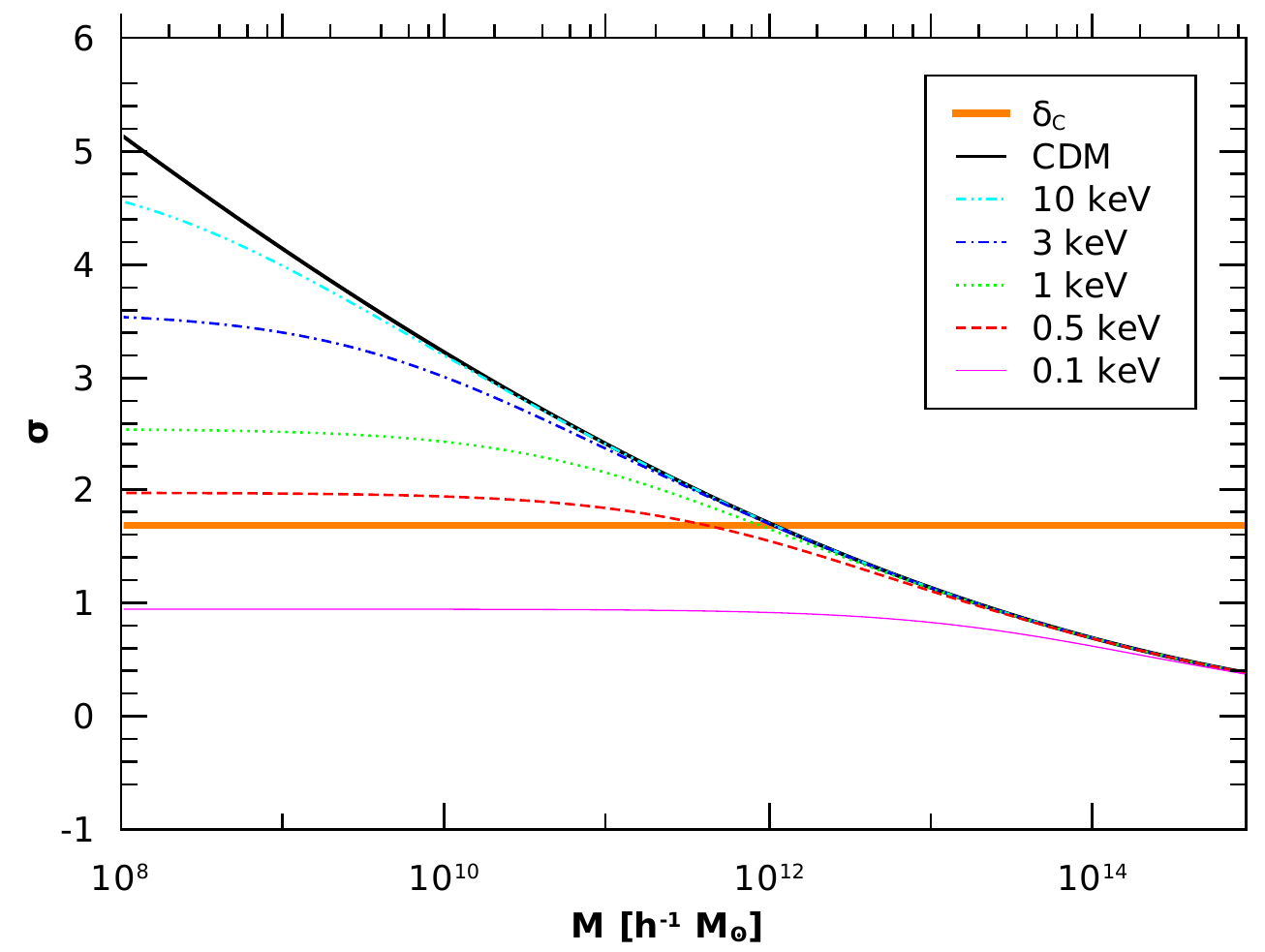}
\caption{\label{img:sigma}Shown is $\sigma\left(M\right)$ for the CDM
  model and several WDM particle masses, as given in the legend.
  Here, $M$ satisfying $\sigma\left(M\right)=\delta_\text{c}$ is the
  typical mass of halos which are currently collapsing.  As shown, for
  very small WDM particle masses, such as 0.1 keV, typical
  fluctuation-scale halos have not collapsed, and little structure would
  have formed today.}
\end{figure}

In the case of WDM, $M_\star$ is not well-defined for all particle
masses.  For light masses, $\sigma\left(M\right)$ is less than
$\delta_\text{c}$ for all masses (see Fig.~\ref{img:sigma}).  This
means that there are no typical halos that are currently collapsing.
Since that is certainly not the case in the universe, these particle
masses of WDM are not realistic.

The halo bias calculated from $\sigma\left(M\right)$ for the CDM
model and several WDM particle masses is shown in Fig.~\ref{img:bias}.
The bias is defined in terms of $\nu$ which, in turn, depends on
$\sigma\left(M\right)$.  Therefore, changes to the linear matter
power spectrum also affect the bias function.  Since the halo bias
function has not been studied in WDM structure simulations, we use the
bias function from Ref.~\cite{Tinker:2010my} for both the CDM and WDM
models.  Though the halo bias should be tested in the case of WDM with
simulations, it is beyond the scope of the present work.

\section{\label{haloMF}Halo Mass Function}

\subsection{\label{haloMF:CDM}Cold Dark Matter}
%Taken from~\cite{Tinker:2008ff}
The halo mass function, $dn/dM$ is the comoving number density of
halos as a function of halo mass $M$ and redshift $z$.  From the
spherical collapse model, overdensities in the matter density field
must be at or above the critical density $\delta_\text{c}$ to
collapse.  (Recall the definition of $M_\star$ in equation
\eqref{eqn:Mstar}.)  The halo mass function can be estimated from the
number of overdensities above the critical density.  The following
form was developed for the mass function, given Gaussian fluctuations
in the initial density field:
\begin{equation}
  \frac{m^2}{\bar{\rho}_z}\frac{dn}{dM}\left(M,z\right)\frac{dM}{M}=\nu
  f\left(\nu\right)\frac{d\nu}{\nu}. \end{equation} The variable
$\nu$ is defined in Eq.~\eqref{eqn:nu}, and $\bar{\rho}_z$ is the
average density of matter at redshift $z$.  The function
$f\left(\nu\right)$ is defined by
\begin{equation} \nu f\left(\nu\right) = \sqrt{\frac{\nu}{2\pi}}\exp{\left(-\frac{\nu}{2}\right)}. \end{equation}

The PS halo mass function is an approximation.  Motivated by
extensions to PS theory, numerical simulations find similar but more
accurate relations for the halo mass function. We use the halo mass
function from Ref.~\cite{Tinker:2008ff}, Appendix C:
\begin{equation}\label{eqn:dn_T} 
\frac{dn}{dM} =
g\left(\sigma\right)\frac{\bar{\rho_0}}{M}\frac{d\ln\sigma^{-1}}{dM} 
\end{equation}
with $g\left(\sigma\right)$ defined as
\begin{equation} 
  \label{eqn:fsig} g\left(\sigma\right) =
  B\left(\left(\frac{\sigma}{e}\right)^{-d}+\sigma^{-f}\right) \exp\left(-\frac{g}{\sigma^2}\right). 
\end{equation}
Their results for the constants $B$, $d$, $e$, $f$ and $g$
at several values of the mean interior density $\Delta$ are given in
Ref.~\cite{Tinker:2008ff}, Appendix C.

\subsection{\label{haloMF:WDM}Warm Dark Matter}

\begin{figure}
\includegraphics[width=\linewidth]{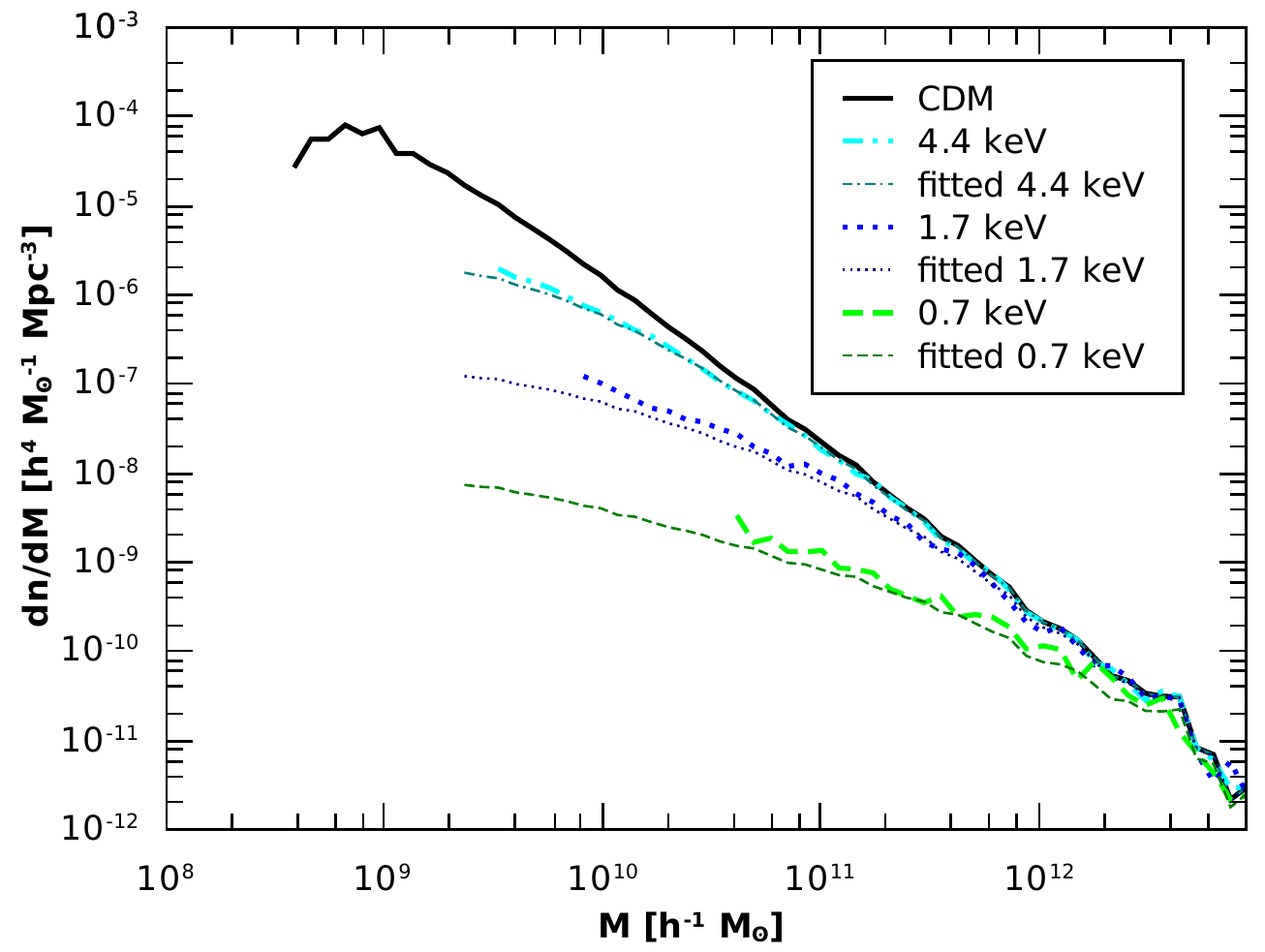}
\caption{\label{img:dndm_fit}Shown is the halo mass function from our
  simulations and our fits to that data.  The thick lines are the data
  from each simulation, as listed in the legend.  The thin lines are
  the CDM data multiplied by our fitting factor, which is given in
  Eq.~\eqref{eqn:dn_WDM}.}
\end{figure}

We find the halo mass function for CDM and three WDM particle masses
in our simulations, described in \S \ref{sims}.  Our data shows that
the suppression of the power spectrum on small scales, as expected,
results in fewer low mass halos.  We removed the upturn in number of
halos with the smallest masses, which is a numerical artifact due to
the discrete nature of the simulation~\cite{Wang:2007he}.  We fit the
WDM mass function as
\begin{equation}
\label{eqn:dn_WDM} 
\frac{dn'_\text{W}}{dM} =
\left(1+\frac{M_\text{f}}{M}\right)^{-\eta}\frac{dn_\text{C}}{dM}, 
\end{equation}
which does not include the erasure of the smallest scale halos. Here,
$dn_\text{C}/dM$ is the CDM mass function, and $M_\text{f}$ is the
filtering mass, which is defined as
\begin{equation} 
\label{eqn:Mf}
M_\text{f}=\frac{4\pi}{3}\bar{\rho}_0\frac{\pi^3}{k_\text{f}^3} 
\end{equation}
where $k_\text{f}$ is the wavenumber at which the WDM transfer
function, Eq.~\eqref{eqn:T_wdm}, has an amplitude of 1/2. For sterile
neutrinos with masses 0.5 keV, 1 keV, 3 keV and 10 keV, $k_f \approx 1.6$,
$2.8$, $7.3$, and $20.5\ h\rm Mpc^{-1}$, respectively, and the filtering
masses are $2.1\times10^{12}\ h^{-1}M_\odot$, $3.6\times10^{11}\
h^{-1}M_\odot$, $2.1\times10^{10}\ h^{-1}M_\odot$ and $9.8\times10^8\
h^{-1} M_\odot$ respectively.  Eq.~\eqref{eqn:dn_WDM} is an accurate
fit to the simulation's mass function for $\eta = 1.2$ (see
Fig.~\ref{img:dndm_fit}).

Since the WDM particles free-stream out of the small scale
perturbations, dark matter halo formation is strongly suppressed below
the free-streaming scale $R_\text{fs}$, which is defined as
\begin{equation} 
  \label{eqn:Rf}
R_\text{fs} = 0.2 \left(\Omega_\text{dm} h^2\right)^{1/3} m_\text{s}^{-4/3}.
\end{equation}
This corresponds to a free-streaming scale mass $M_\text{fs}$ at
mean density (not to be confused with the filtering mass
$M_\text{f}$). We use this mass for our small scale cut-off in the
mass function.  We use an expression for the small halo suppression
effect that is approximately unity when $m \gg M_\text{fs}$ and goes
to zero when $m \ll M_\text{fs}$ to provide a continuous
cut-off. Our WDM mass function is then
\begin{equation}\label{eqn:dn_WDM_2}
  \frac{dn_\text{W}}{dM}=\left(\frac{M^2}{M^2+M_\text{fs}^2}\right)^{10^3}
  \left(1+\frac{M_\text{f}}{M}\right)^{-\eta}\frac{dn_\text{C}}{dM}.
\end{equation}

\begin{figure}
\includegraphics[width=\linewidth]{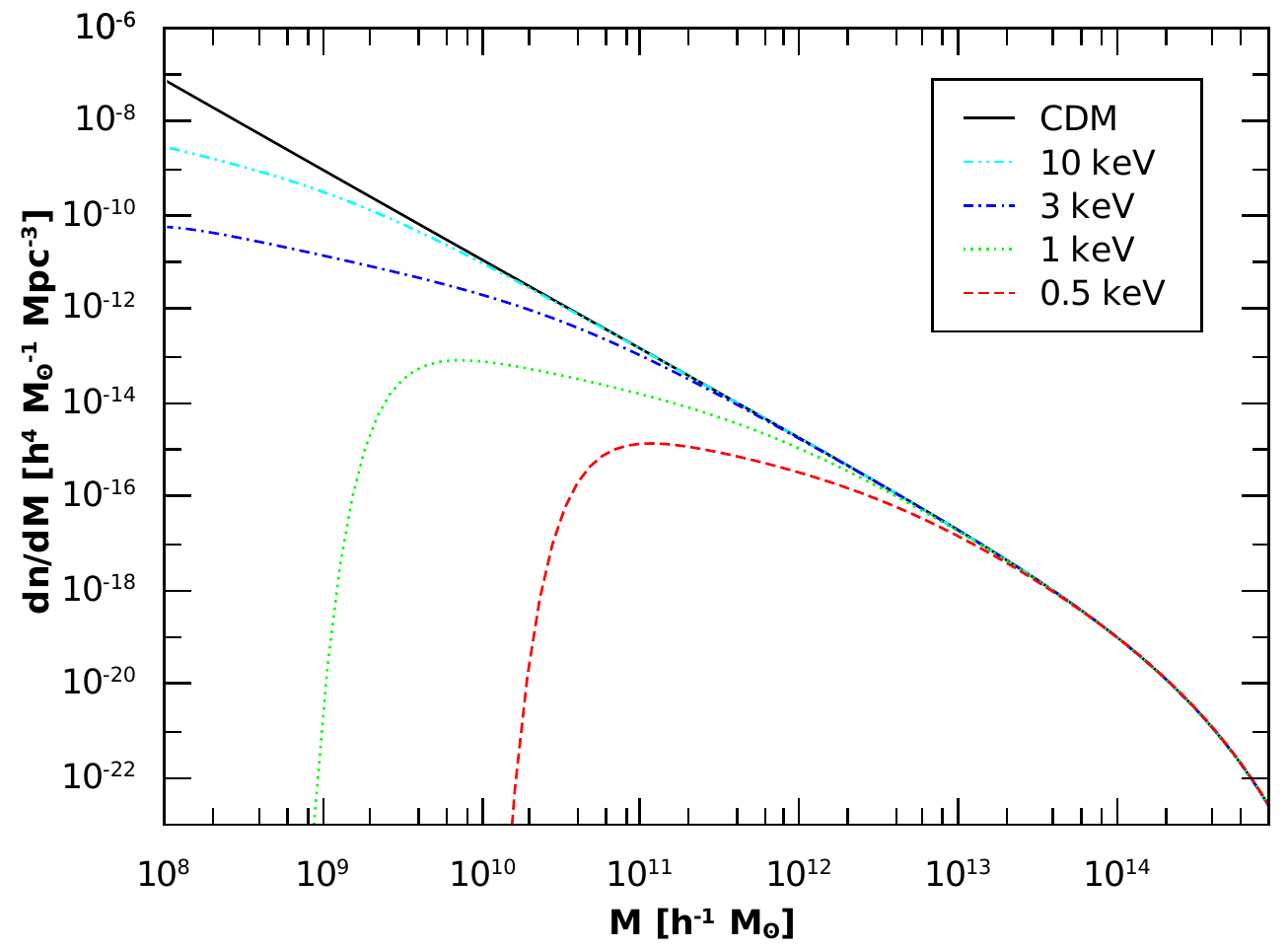}
\caption{\label{img:dndM}Shown is the halo mass function for the CDM
  model and several WDM particle masses, as given in the legend.  The
  mass functions are normalized to the CDM mass functions at large
  masses.  Our WDM mass functions include a suppression for small
  $M$ and a small scale cut-off.}
\end{figure}

\begin{figure}
\includegraphics[width=\linewidth]{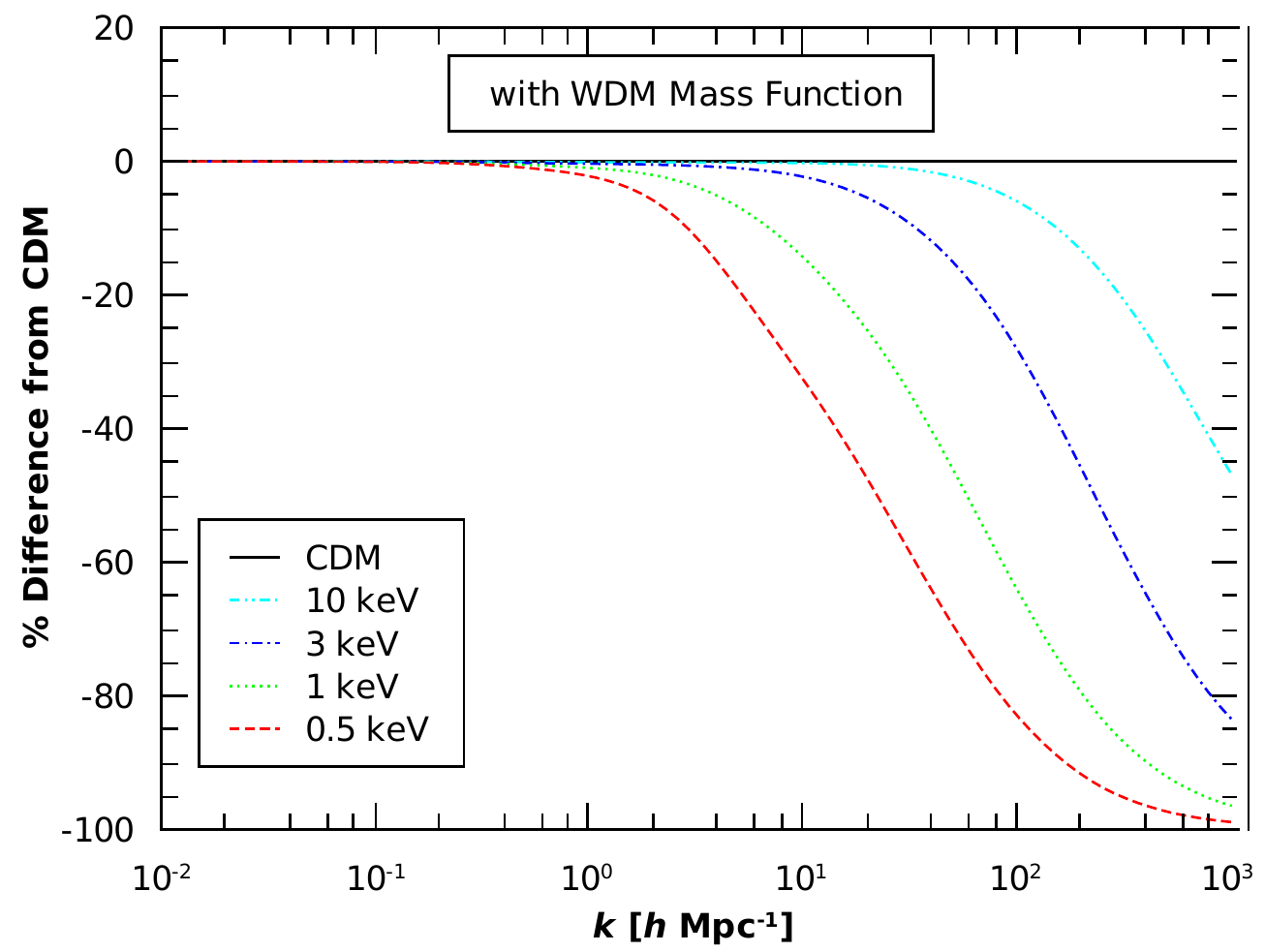}
\caption{\label{img:Pnl_8}The percent difference between the nonlinear
  matter power spectrum for the CDM model and several WDM particle
  masses, as listed in the legend, is plotted versus wavenumber $k$.
  In this graph, the only difference between the CDM and WDM models is
  the change to the mass function, as given in Eq.s
  \eqref{eqn:dn_WDM_2}.  The decrease in small mass halos results in
  the suppression of the power spectrum at small scales.}
\end{figure}

The full mass function for our CDM and WDM models is shown in
Fig.~\ref{img:dndM}.  The effect of just the change to the mass
function on the nonlinear matter power spectrum is shown in
Fig.~\ref{img:Pnl_8}.  The decrease in small mass halos results in the
suppression of the power spectrum at small scales.

\section{\label{ukm}Halo Density Profile}

\subsection{\label{ukm:CDM}Cold Dark Matter}
A general halo density profile is given by Ref.~\cite{Colin:2007bk}:
\begin{equation} 
\label{eqn:rho} 
\rho\left(r\mid M\right) =
\frac{\rho_\text{s}}{\left(\frac{r}{r_\text{s}}\right)^\gamma
\left[1+\left(\frac{r}{r_\text{s}}\right)^\alpha\right]^{(\beta-\gamma)/\alpha}}.
\end{equation}

For CDM, we employ the Navarro, Frenk and White (NFW)
profile~\cite{NFW96}, which has $\alpha=1$, $\beta=3$ and
$\gamma=1$.  The radius $r_\text{s}$ is the radius at which the
density function has a logarithmic slope of $-2$.  For a spherically
symmetric density profile, the mass of the halo is
\begin{equation} 
M\equiv\int_0^{R_\text{vir}}dr4\pi r^2\rho\left(r\mid M\right), 
\end{equation}
for the virial radius $R_\text{vir}$.  For the NFW profile, the
integral has an analytic solution, which specifies the value
of~$\rho_\text{s}$:
\begin{equation} 
M=4\pi\rho_\text{s}
  r_\text{s}^3\left(\ln\left(1+c\right)-\frac{c}{1+c}\right). 
\end{equation}
The concentration $c$ is defined as $c \equiv
R_\text{vir}/r_\text{s}$.  When calculating the nonlinear matter
power spectrum, we use the Fourier transform of the density profile,
\begin{equation} 
  u\left(k\mid M\right) = \int_0^{R_\text{vir}}dr4\pi
  r^2\frac{\sin\left(kr\right)}{kr}\frac{\rho\left(r\mid
      M\right)}{M}. 
\end{equation}

\subsection{\label{ukm:WDM}Warm Dark Matter}

The density profile of halos in WDM simulations has previously been
studied by Avila-Reese and Col\'{i}n et al.~\cite{Colin:2000dn,
  AvilaReese:2000hg, Colin:2007bk}.  In~\cite{Colin:2000dn}, Milky Way
size halos are simulated for WDM particles of mass 0.6, 1. and
2. keV. Ref.~\cite{Colin:2000dn} found that the density profile of the
halos is described by the NFW profile, but WDM halos can have an inner
slope slightly shallower than $-1$.  In Ref.~\cite{AvilaReese:2000hg},
they simulate CDM and WDM halos with masses down to $~0.01M_\text{f}$.
These halos were found to also be well described by the NFW profile.
For WDM halos with masses below $M_\text{f}$, it was found that the
inner slope is shallower than in comparable CDM halos.  In
Ref.~\cite{Colin:2007bk}, halos with masses close to the filtering
mass $M_\text{f}$ were simulated, which found that the innermost ($r
\lesssim 0.02R_\text{vir}$) logarithmic slope of WDM halos is steeper
than in the CDM model.  Outside of this volume, the density profile
for WDM halos is shallower than the NFW fit.  Ref.~\cite{Colin:2007bk}
finds that, using the density profile of Eq.~\eqref{eqn:rho}, a cored
profile with $\alpha=0.7$, $\beta=3$ and $\gamma=0$ fits the WDM
halos.  Ref.~\cite{Smith:2011ev} uses a halo density profile where the
size of the core increases as the mass of the halo decreases.  The
small mass halos $\left(M \lesssim 10^{11}\ h^{-1} M_\odot\right)$, in
Ref.~\cite{Smith:2011ev} have cores with radii on the order of
$r_\text{s}$.  On the other hand, more recent work in
Refs.~\cite{deNaray:2009xj,VillaescusaNavarro:2010qy} studies whether
WDM halos can have such cores, and shows that, in halo collapse
modeling, the core of a WDM halo is smaller than $r \lesssim 10^{-3}
R_\text{vir}$.  Below this scale, a core may need to be included in
the density profile.  Therefore, it is unlikely that WDM halos have
large cores.

WDM halos that are smaller than the filtering mass (see
Eq.~\eqref{eqn:Mf}) have a shallower inner slope compared to CDM
halos~\cite{Colin:2000dn, AvilaReese:2000hg}.
Ref.~\cite{Ricotti:2002qu} shows that, in CDM simulations, the slope
of the inner profile depends on the effective spectral index
$n_\text{eff}$ of the initial power spectrum of the density
perturbations at the scale of the power spectrum sampled by the halo
$k_\text{halo}$.  That is, $P\left(k_\text{halo}\right) \propto
k_\text{halo}^{n_\text{eff}}$.  Ref.~\cite{Ricotti:2002qu} finds that
\begin{equation} 
\gamma = \frac{9+3n_{\rm eff}}{5+n_{\rm eff}}, 
\end{equation}
where $\gamma$ is an exponent from Eq.~\eqref{eqn:rho}.  This is
relevant to our WDM model since the suppression of the small-scale
structure acts to decrease the effective spectral index at small
scales, as in Fig.~\ref{img:Plin}.  This is what leads to a decrease
in $\gamma$, the inner slope of the profile, though since $n_{\rm
  eff}$ is not well defined in WDM for a fixed halo mass, one cannot
specify $\gamma$ analytically.

\begin{figure}
\includegraphics[width=\linewidth]{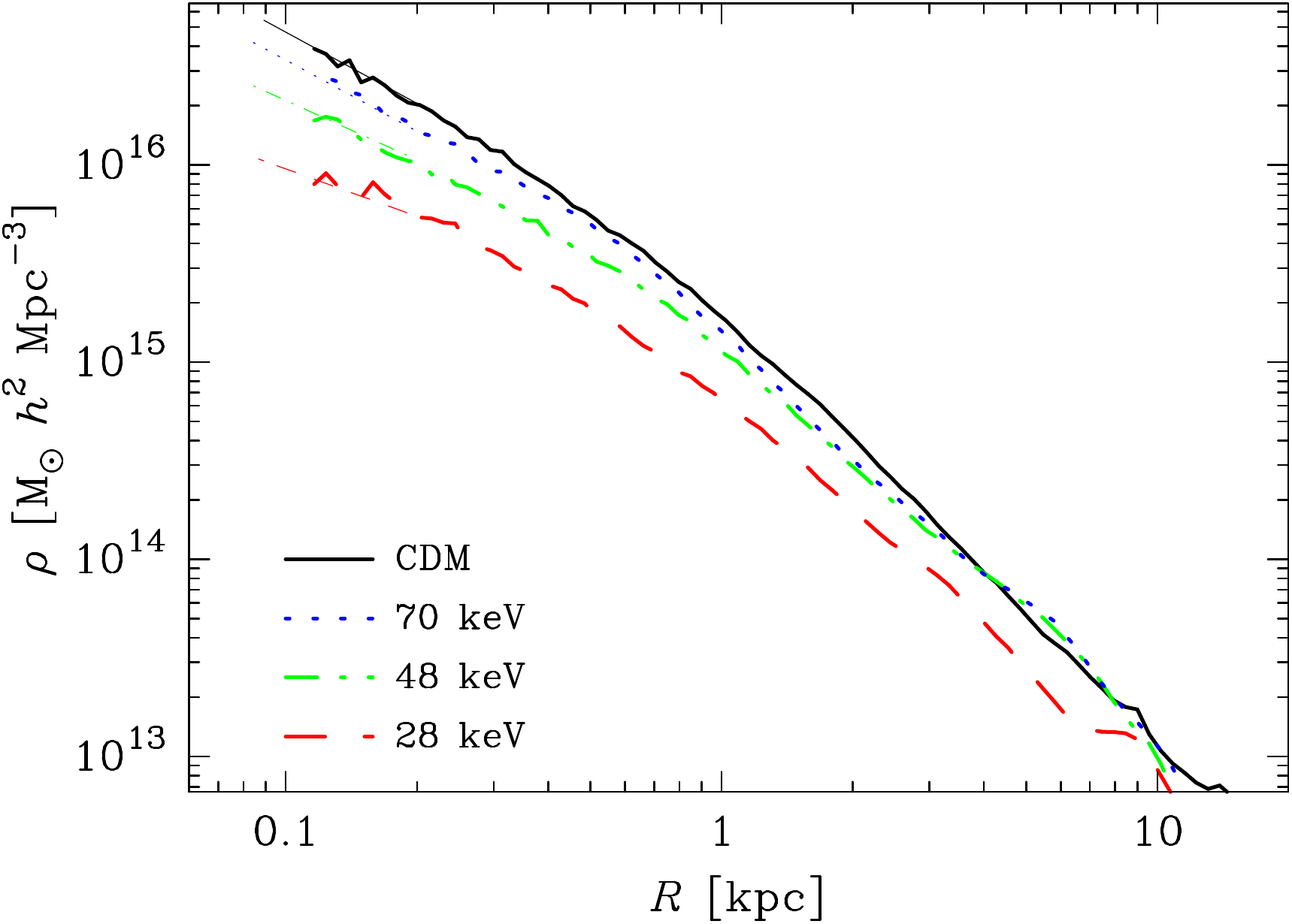}
\caption{\label{img:rho_data}This is a plot of the density versus
  radius for a $10^8 M_\odot$ halo simulated in four different WDM
  cosmologies, as given in the legend.  The thick lines are data from
  the simulation, and thin lines show the estimated inner slope for
  the halo density profile, with $\gamma=1.2$, $\gamma=1.2$,
  $\gamma=1.0$, $\gamma=0.8$ for the CDM, 70 keV, 48 keV and 28 keV
  cases, respectively.  The value $\gamma$ describes the profile inner
  slope in Eq.~\eqref{eqn:rho}.}
\end{figure}

We use our simulation of a 10$^8$ $M_\odot$ halo for CDM and three
sterile neutrino WDM particle masses: 28 keV, 48 keV, and 70 keV.  The
density profile of these halos are shown in Fig.~\ref{img:rho_data}.
We find that the simulated CDM and 70 keV WDM halos have an inner slope
of $-1.2$.  The inner slope of the simulated 48 keV WDM halo is $-1.0$,
and the simulated 28 keV WDM halo has an inner slope of $-0.8$.  We use
an inner slope of $-1$ for the CDM cosmology, as in the NFW profile.  To
model the effect of the suppression of the density profile, the inner
slopes of halos composed of 70 keV, 48 keV, and 28 keV WDM particles are
chosen to be $-1$, $-0.8$ and $-0.6$, respectively.  We use these
values to interpolate the inner slope for other WDM particle masses.

To include this effect, we make changes to the exponents in the
profile density function (Eq.~\eqref{eqn:rho}).  The exponent $\gamma$
describes the logarithmic slope of the central section of the halo,
where $r/r_\text{s} \ll 1$.  For the inner slope to be shallower, we
use $\gamma < 1$.  The smaller the mass of the WDM particle, the
shallower the inner slope.  When the inner slope of the density
profile decreases, the density of the central region of the halo
increases for a constant halo mass.

The exponent $\beta$ describes the slope of the density profile for
the outer edge of the halo, where $r/r_\text{s} \gg 1$.  The quantity
$\alpha$ describes the sharpness of the change between the inner and
outer sections of the density profile.  For larger $\alpha$, the
changeover between a slope of $\gamma$ and a slope of $\beta$ is
sharper.  It takes place within a smaller region around
$r/r_\text{s}=1$.

To maintain the definition of $r_\text{s}$ as the radius at which
the density profile has a logarithmic slope of $-2$, the exponents
are constrained to be:
\begin{equation} 
  1=\sqrt[\alpha]{\frac{\gamma-2}{2-\beta}} \Rightarrow
  \beta = 4-\gamma .
\label{eqn:dens_exp} 
\end{equation} Note that this
relation is valid for $\alpha = 1$, $\beta = 3$ and $\gamma =
1$, which is the NFW profile.  

\begin{figure}
\includegraphics[width=\linewidth]{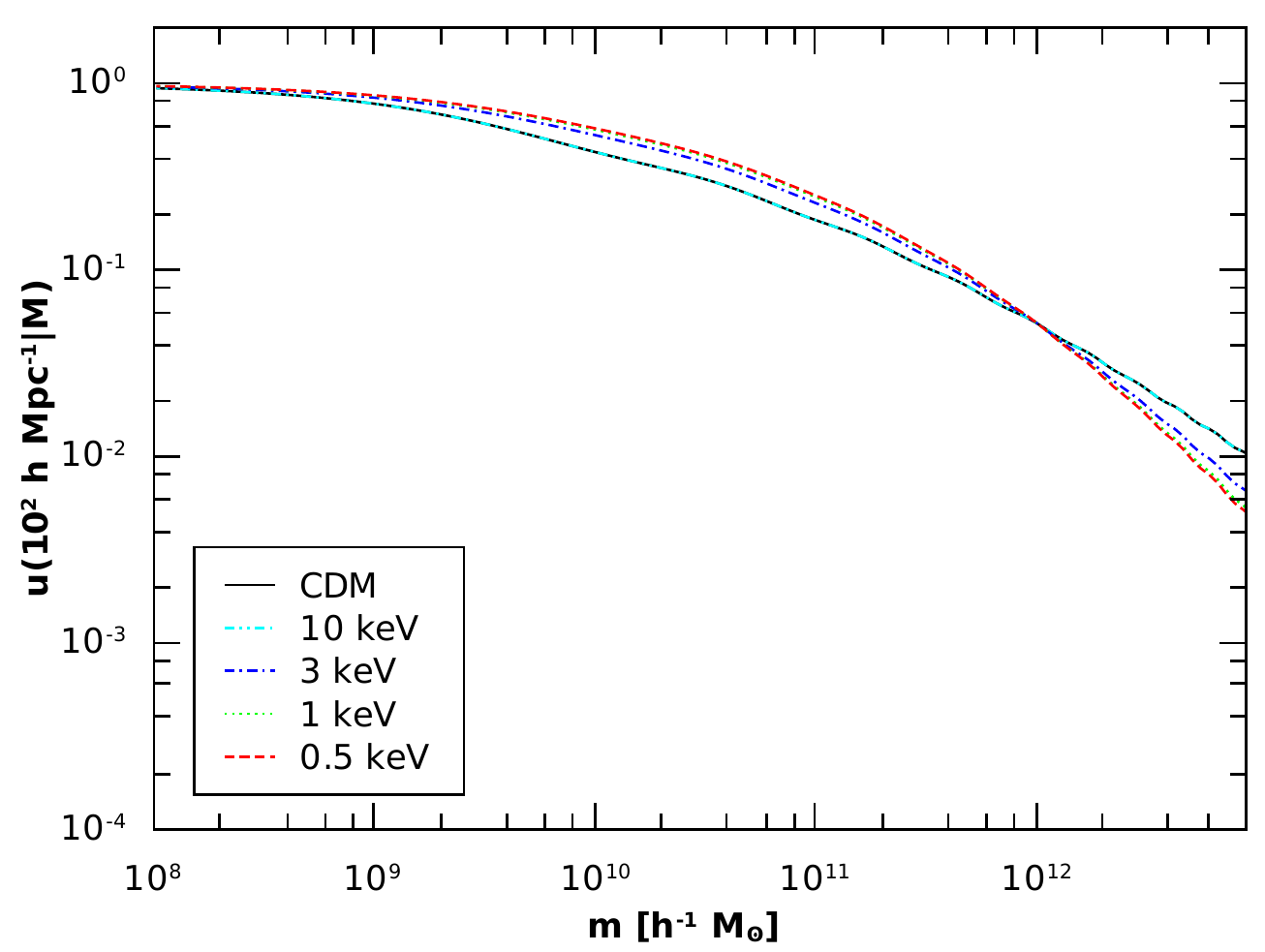}
\caption{\label{img:uM}The Fourier transform of the halo density
  profile $u\left(k|M\right)$ for a constant wavenumber of $10^2\
  h^{-1} \text{Mpc}$ is plotted as a function of halo mass $M$ for
  the models listed in the legend.}
\end{figure}

\begin{figure}
\includegraphics[width=\linewidth]{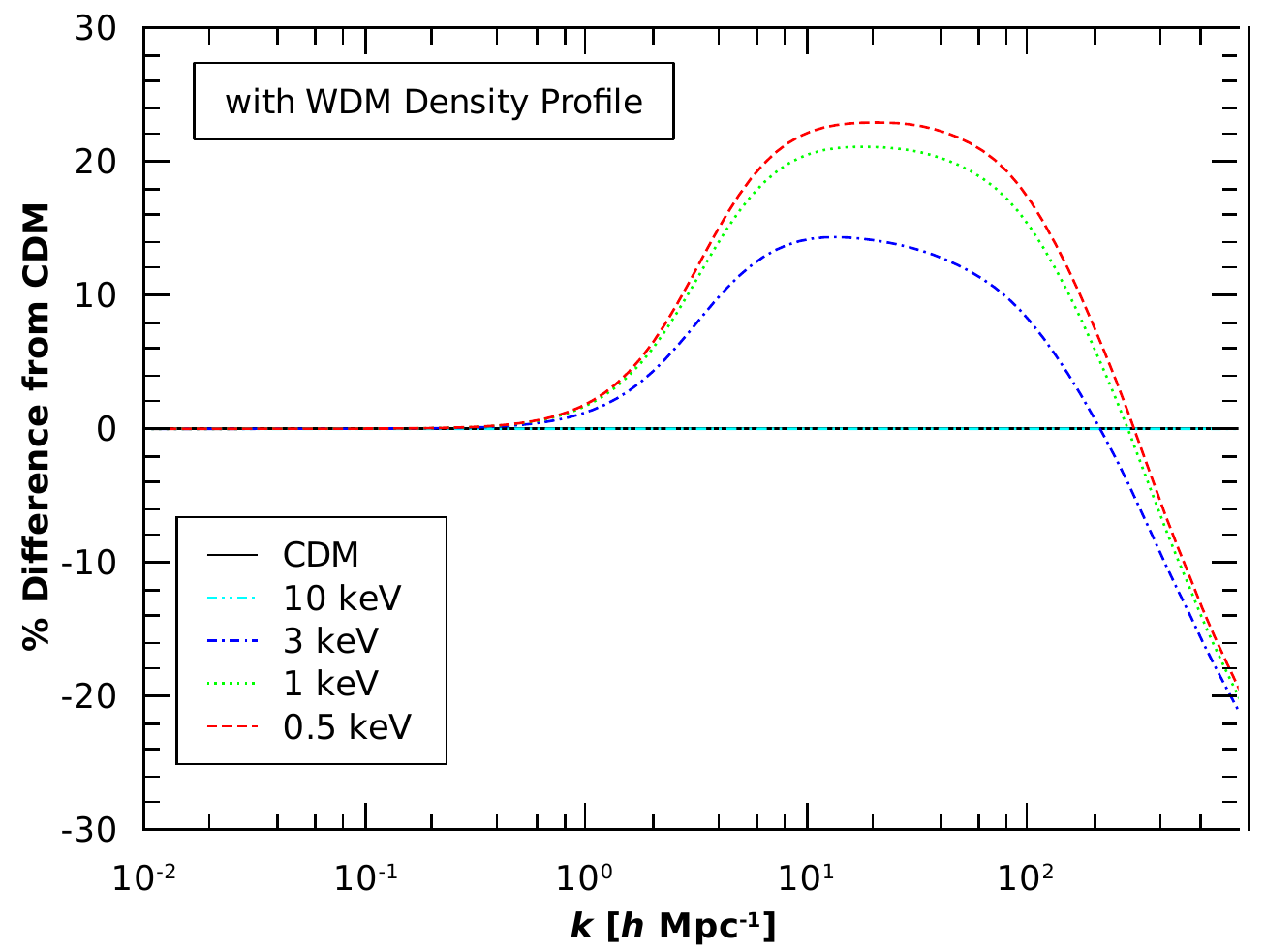}
\caption{\label{img:Pnl_4}The percent difference between the nonlinear
  matter power spectrum for the CDM model and several WDM particle
  masses, as listed in the legend, is plotted versus wavenumber $k$.
  In this graph, the only difference between the CDM and WDM models is
  the change to the inner slope of the halo density profile.  The
  broadening of the central part of the halo and suppression of the
  peak results in a transfer of power
  from the smallest scales to slightly larger scales, as shown by the
  peak in the above plot.}
\end{figure}

Since the profile exponent change in $\gamma$ changes the density
distribution, in order to maintain a constant halo mass, the value of
$\rho_s$ must be rescaled with the profile changes, which we
incorporate in the WDM models.  Shallower $\gamma$ requires an
increase in $\rho_s$, which broadens the density profile for a fixed
halo mass.  The Fourier transform of the density profile is employed
in the nonlinear power spectrum.  Fig. \ref{img:uM} is a plot of
$u\left(k|M\right)$ for a constant wavenumber $k$.  The increase in
density around the halo center increases $u\left(k|M\right)$ for large
$k$ given a constant $M$.  Fig.~\ref{img:Pnl_4} shows the effect of
changes to the halo density profile on the nonlinear matter power
spectrum.  The broadening of the central part of the halo and
suppression of the central peak profile results in a transfer of power
from the smallest scales to slightly larger scales, seen in
Fig.~\ref{img:Pnl_4}.

\section{\label{concen}Concentration}

The concentration of an NFW (see \S \ref{ukm}) CDM halo has been found
to be~\cite{Bullock:1999he}
\begin{equation} 
\bar{c}\left(m,z\right) \simeq
\frac{9}{1+z}\left(\frac{m}{m_\star\left(z\right)}\right)^{-0.13}.
\label{eqn:concCDM} 
\end{equation}
We apply the methodology described in Ref.~\cite{Bullock:1999he} to
the WDM linear power spectrum to determine the change in halo
concentration.

The model in Ref.~\cite{Bullock:1999he} is determined by two equations
and the parameters $F$ and $K$:
\begin{equation}
  \label{eqn:conc_F} 
M_\star\left(z_\text{c}\right)\equiv FM, 
\end{equation}
and
\begin{equation}
  c\left(M,z_\text{c}\right)=K\frac{1+z_\text{c}}{1+z}, 
\end{equation}
where $M_\star$ is defined in Eq.~\eqref{eqn:Mstar}.

\begin{figure}
  \includegraphics[width=\linewidth]{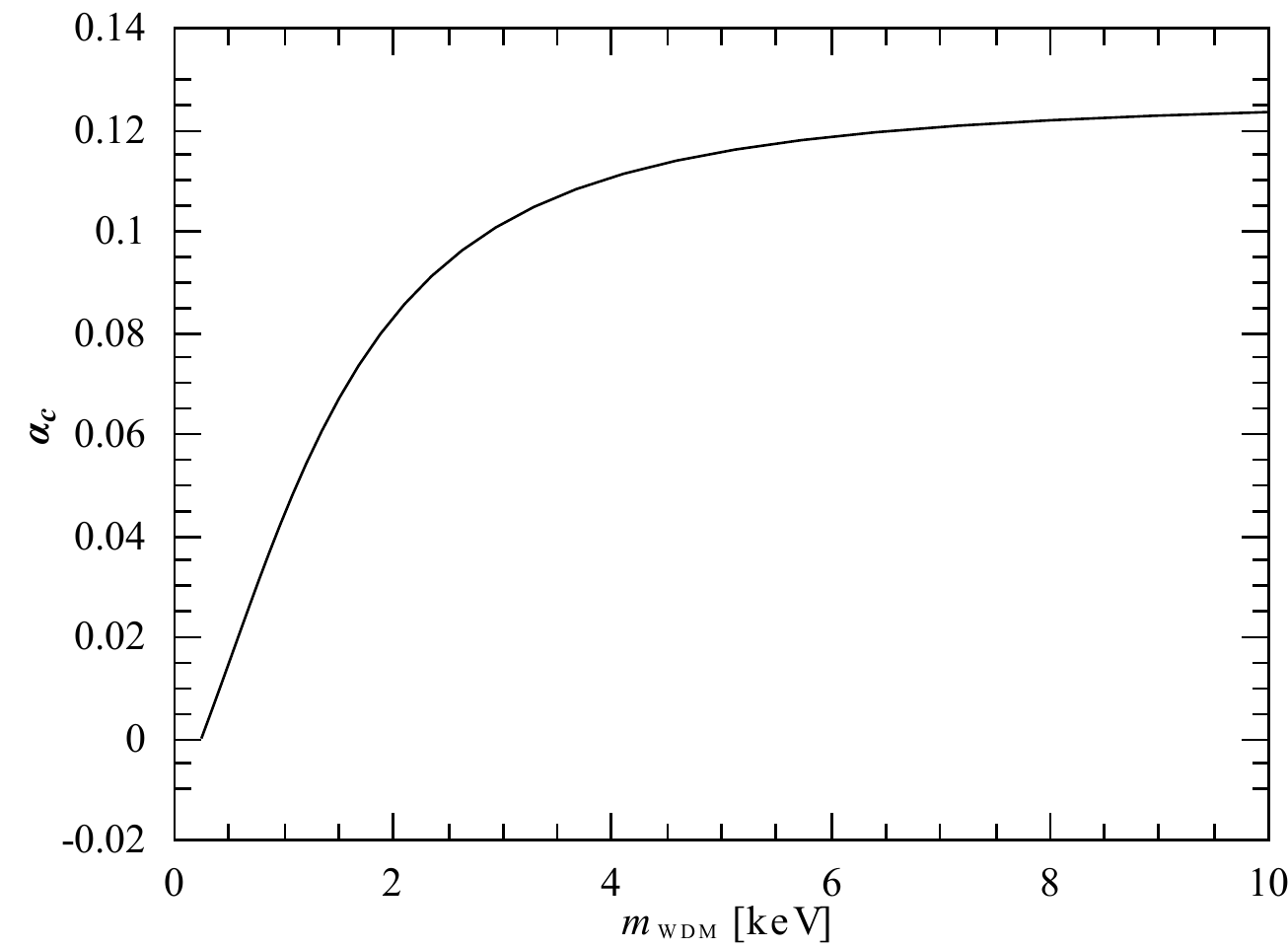}
  \caption{\label{img:alpha_conc}The exponent in the concentration
    equation, Eq.~\eqref{eqn:conc_alpha}, is plotted as a function of
    the WDM particle mass.  As particle mass increases, $\alpha_c$
    approaches 0.13, its value in the CDM model.}
\end{figure}

\begin{figure}
\includegraphics[width=\linewidth]{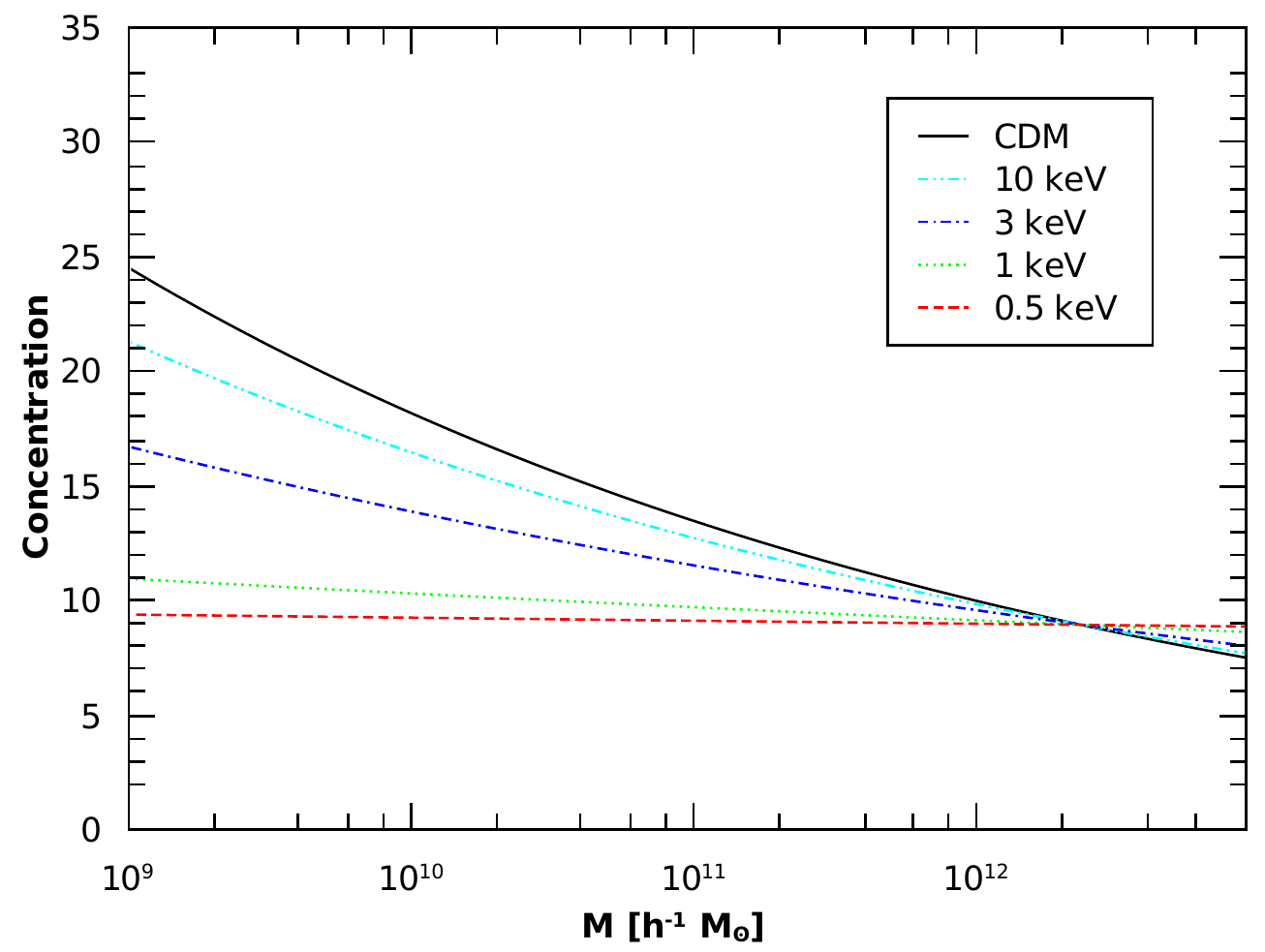}
\caption{\label{img:conc}The concentration of halos is plotted as a
  function of halo mass for the CDM model and several WDM particle
  masses, as given in the legend.  Since the exponent $\alpha_c$
  decreases with WDM particle mass (see Fig.~\ref{img:alpha_conc}),
  the dependence of the concentration on halo mass also decreases.}
\end{figure}

We assume that $\sigma\left(M\right)$ is approximately a power law
at a mass of $FM_\star$:
\begin{equation}\label{eqn:conc_alpha} 
  c \propto M^{-\alpha_c}. 
\end{equation}
Then, we can calculate $\alpha_c$ from $\sigma\left(M\right)$:
\begin{equation} 
\alpha_c =
  \frac{d\left(\ln\sigma^{-1}\right)}{d\left(\ln M\right)} =
  \frac{-M}{\sigma}\frac{d\sigma}{dM}. 
\end{equation} 
For the CDM case, the best fit parameters for $F$ and$K$ are 0.01 and
4.0, respectively.  Then, $\alpha_c = 0.13$, as in
Eq.~\eqref{eqn:concCDM}.  This is valid for $0.01 M_\star \lesssim M
\lesssim 100 M_\star$~\cite{Bullock:1999he}.
Fig.~\ref{img:alpha_conc} shows $\alpha_c$ as a function of WDM
particle mass.  The resulting concentration for halos in the CDM
models and for several WDM masses is shown in Fig.~\ref{img:conc}.
Since the exponent $\alpha_c$ decreases with WDM particle mass (see
Fig.~\ref{img:alpha_conc}), the dependence of the concentration on
halo mass also decreases.  In addition, the concentration of halos in
WDM models is less than in CDM~\cite{Colin:2007bk, AvilaReese:2000hg,
  Colin:2000dn}.
%Look up papers and cite that say that the concentration is less for WDM

\begin{figure}
\includegraphics[width=\linewidth]{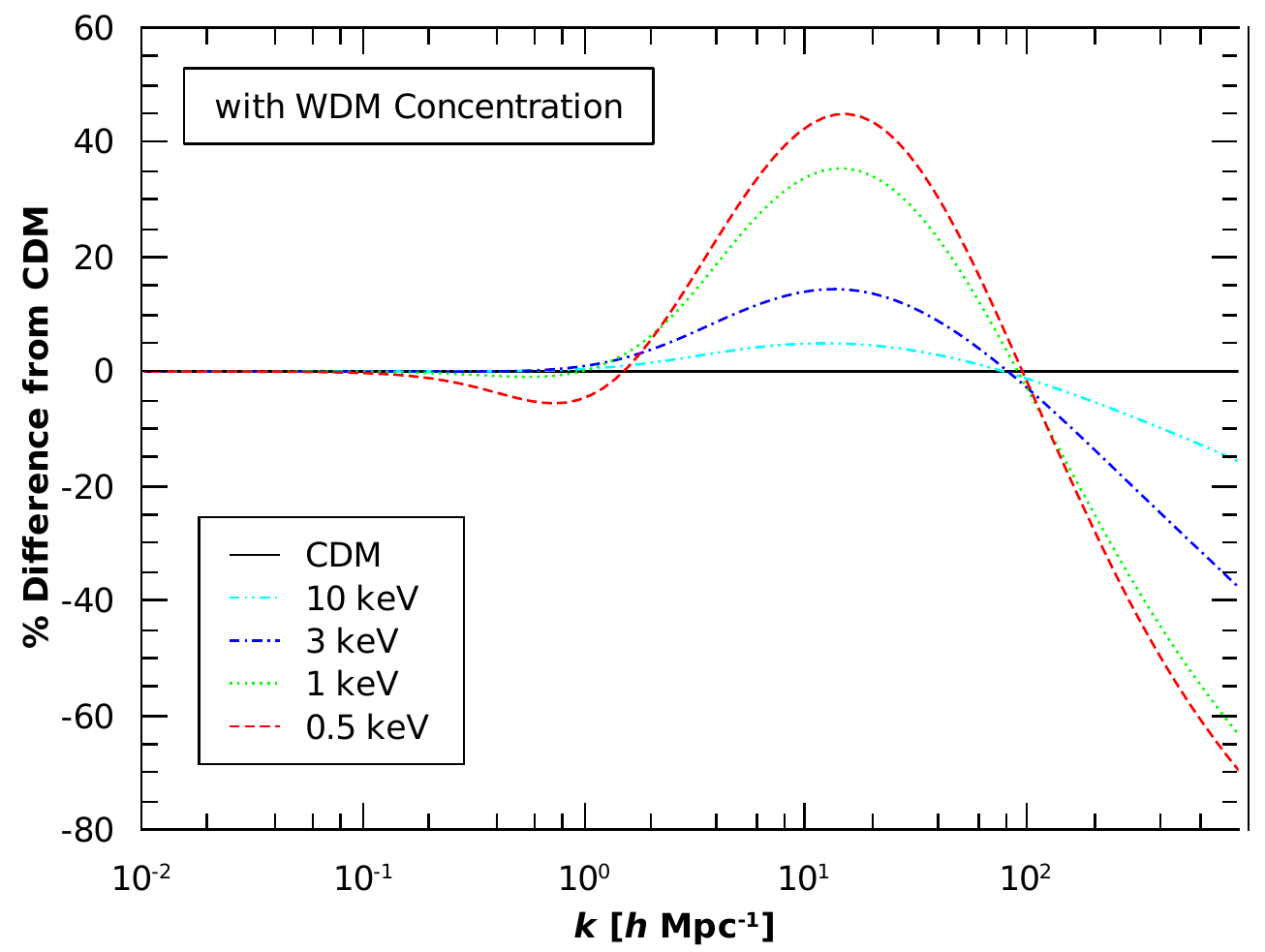}
\caption{\label{img:Pnl_3}The percent difference between the nonlinear
  matter power spectrum for the CDM model and several WDM particle
  masses, as listed in the legend, is plotted versus wavenumber $k$.
  In this graph, the WDM models differ from the CDM model in both the
  transfer function and concentration.  The decrease in the
  concentration (see Fig.~\ref{img:conc}) increases the scale radius
  $r_\text{s}$, from Eq.~\eqref{eqn:rho}, which stretches the inner
  part of the halo profile relative to the radius.  This results in a
  transfer of power from the smallest scales to slightly larger
  scales, as shown by the peak in the above plot.}
\end{figure}

Since $\alpha_c$ depends on the linear matter power spectrum through
$\sigma\left(M\right)$, changing the concentration has an effect
only if the linear matter power spectrum also changes.
Fig.~\ref{img:Pnl_3} shows the effect of the WDM linear matter power
spectrum and concentration on the nonlinear matter power spectrum.
The decrease in the concentration (see Fig.~\ref{img:conc}) increases
the scale radius $r_\text{s}$, from Eq.~\eqref{eqn:rho}, which
stretches the inner part of the halo profile relative to the radius.
This also results in a transfer of power from the smallest scales to
slightly larger scales, as shown by the peak in the nonlinear matter
power spectrum.

\section{\label{sub}Substructure}

% The changes to the 1-halo term to add substructure are taken
% from~\cite{Dolney:2004cn}.  check concentration for subhalos to
% correct in WDM
Clearly, in WDM, small mass halos are suppressed in number, and
therefore subhalos will be suppressed as they are accreted halos in
the hierarchical formation of large scale structure.  Dolney et
al.~\cite{Dolney:2004cn} study the effects of substructure on the CDM
halo model.  A halo consists of a smooth mass component and several
subhalos.  Most of the mass of the halo is in the smooth component.  A
fraction $f$ of the mass is in the subhalos.
Ref.~\cite{Dolney:2004cn} uses $f\approx0.1$ and assumes that all
subhalos are smaller than $0.01M$ where $M$ is the mass of the
host halo.  With substructure, the one halo term becomes a sum of four
different terms:
\begin{equation}
\label{eqn:1hs} 
P_\text{1h}=P_\text{ss}+P_\text{sc}+P_\text{1c}+P_\text{2c}. 
\end{equation}
The first term in Eq.~\eqref{eqn:1hs}, $P_\text{ss}$ describes
correlations between two points of the smooth component of the host
halo:
\begin{equation} 
  P_\text{ss}\left(k\right)=\int
  dM\frac{dN}{dM}\left(\frac{M_\text{s}}{\bar{\rho}}\right)^2\left|U\left(k\mid
      M_\text{s}\right)\right|^2. 
\end{equation} 
Throughout this section, variables that are capital letters describe
the host halo, and lower case variables are associated with the
subhalos.  Note the similarity of Eq.~\eqref{eqn:1hs} to the one halo
term without substructure, Eq.~\eqref{eqn:1h}.  The quantity
$M_\text{s}$ is the mass of the host halo that is in the smooth
component $\left(M_\text{s}=M\left(1-f\right)\right)$, $dN/dM$ is the
mass function for host halos, and $U$ is the density profile for host
halos.  The second term in Eq.~\eqref{eqn:1hs} describes correlations
between the smooth component of the host halo and the subhalos:
\begin{eqnarray}
  P_\text{sc}\left(k\right)&=&2\int dM \frac{dN}{dM}
  \frac{M_\text{s}}{\bar{\rho}}U \left(k\mid M_\text{s}
  \right)U_\text{c} \left(k\mid M_\text{s}\right) \nonumber \\
  &&\times\int dm\frac{dn}{dm}\frac{m}{\bar{\rho}}u \left(k\mid m\right).
\end{eqnarray}
The expression $U_\text{c}$ is the density profile for subhalos
within the host halo, and $dn/dm$ and $u$ are the mass function
and density profile for the subhalos.  The third term from
Eq.~\eqref{eqn:1hs} is the correlation for two points within the same
subhalo:
\begin{eqnarray}
  P_\text{1c}\left(k\right)&=&\int dM\frac{dN}{dM} \nonumber \\
  &&\times\int dm\frac{dn}{dm}\left(\frac{m}{\bar{\rho}}\right)^2\left|u\left(k\mid m\right)\right|^2.
\end{eqnarray}
The fourth term is the correlation between two different subhalos in
the same host halo:
\begin{eqnarray}
  P^\text{2c}\left(k\right)&=&\int dM\frac{dN}{dM}\left|U_\text{c}\left(k\mid M_\text{s}\right)\right|^2 \nonumber \\
  &&\times\left(\int dm\frac{dn}{dm}\frac{m}{\bar{\rho}}u\left(k\mid m\right)\right)^2.
\end{eqnarray}
Dolney et al.~\cite{Dolney:2004cn} assume that the effect of
substructure is negligible on the scale of correlations between two
different halos, and we do the same.  Therefore, the two halo term
(Eq.~\eqref{eqn:2h}) remains the same.

The host halo mass function $dN/dM$ and density profile $U$ are as
defined in the previous sections.  For substructure, we need
expressions for the spatial profile of subhalos within the host halo
$U_\text{c}$, the mass function for subhalos $dn/dm$ and the
density profile with the subhalos $u$.  We use the same density
profile for the subhalos as for the host halos. The distribution of
subhalos in the host halo is taken to be the density profile of the
host halo $U$. The concentration for the subhalos is given
by~\cite{Bullock:1999he}:
\begin{equation}
  \bar{c}\left(m,z\right)=\frac{7.5}{1+z}\left(\frac{m}{M_\star}\right)^{-0.30}. 
\end{equation}
The exponent $-0.30$ corresponds to $F\approx200$ where $F$ is
taken from Eq.~\eqref{eqn:conc_F}.
%We change this concentration with WDM particle mass as described in section \ref{conc}.  
The subhalo mass function is
\begin{equation}
\label{eqn:sub_cdm}
  \frac{dn}{dm}\left(M\right)dm=N_0\left(\frac{M}{m}\right)^\mu\frac{dm}{m}. 
\end{equation}
The constant $N_0$ is determined by the fraction $f$ of the halo
mass in subhalos.  This relation can be solved analytically to find
\begin{equation}
\label{eqn:frac} 
f=\int dm\frac{m}{M}\frac{dn}{dm}\left(M\right) =
\frac{0.01^{\left(1-\mu\right)}N_0}{1-\mu},
\end{equation}
where the most massive subhalos are assumed to be one hundredth of the
mass of the host halos.  Dolney et al.~\cite{Dolney:2004cn} use a
$\mu\approx0.9$ for CDM.

\subsection{Warm Dark Matter}

\begin{figure}
\includegraphics[width=\linewidth]{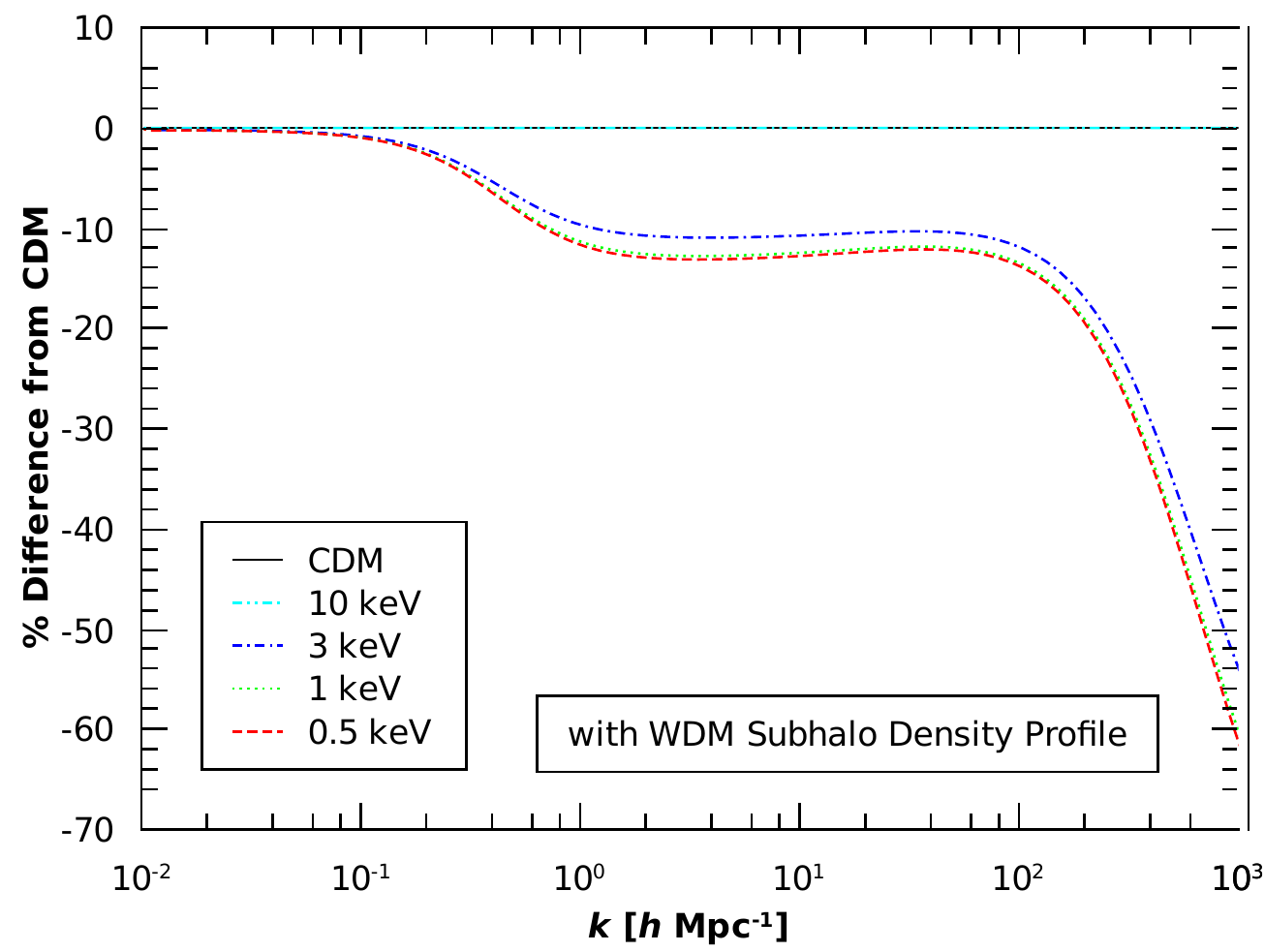}
\caption{\label{img:Pnl_sq}The percent difference between the
  nonlinear matter power spectrum for the CDM model with substructure
  and several WDM particle masses, as listed in the legend, is plotted
  versus wavenumber $k$.  In this graph, the WDM models differ from
  the CDM model in only the subhalo density profile.  As with changes
  to the halo density profile, decreasing the inner slope of the
  subhalo density profile decreases the power on the smallest scales
  (large $k$).}
\end{figure}

In the CDM model, we employ the same density profile for halos and
subhalos.  When considering the effect of WDM on the subhalo density
profile, we alter it in the same was as the halo density profile,
which is described in \S \ref{ukm}.  Fig.~\ref{img:Pnl_sq} shows how
changing the subhalo density profile with WDM particle mass affects
the nonlinear matter power spectrum.  In this figure, the CDM model
also includes substructure.  Since the center of the subhalo clumps
shallower cusps, the smallest scales (highest $k$) transfer power to
slightly larger scales.  But, those scales also lose power to this
effect, so the net result is negative.

\begin{figure}
\includegraphics[width=\linewidth]{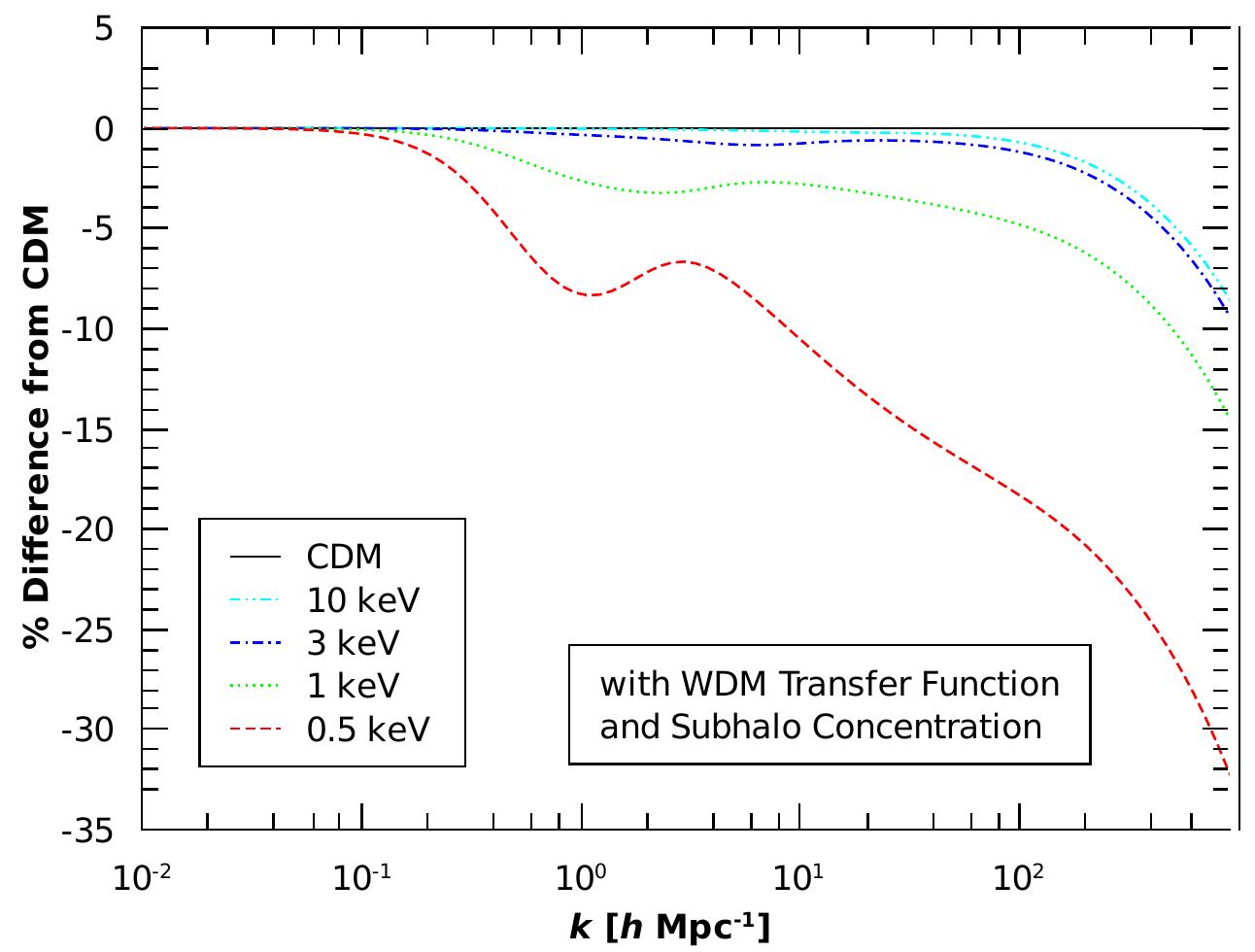}
\caption{\label{img:Pnl_tsb}The percent difference between the
  nonlinear matter power spectrum for the CDM model with substructure
  and several WDM particle masses, as listed in the legend, is plotted
  versus wavenumber $k$.  In this graph, the WDM models differ from
  the CDM model in the transfer function and the subhalo
  concentration.  The concentration of the subhalos decreases, which
  decreases the amplitude of the power spectrum on the smallest scales
  (large $k$).}
\end{figure}

For the subhalo concentration, we follow the procedure outlined in \S
\ref{concen}, but use $F \approx 200$ where $F$ is taken from
Eq.~\eqref{eqn:conc_F}.  This allows us to calculate the exponent
$\alpha_c$, in Eq.~\eqref{eqn:conc_alpha}, for the subhalo
concentration.  For the WDM particle masses 0.5 keV, 1 keV, 3 keV, and
10 keV, this exponent is 0.23, 0.25, 0.26, and 0.26, respectively.
The effect of the subhalo concentration on the nonlinear matter power
spectrum is shown in Fig.~\ref{img:Pnl_tsb}.  The concentration of the
subhalos decreases, which decreases the amplitude of the power
spectrum on the smallest scales (large $k$).

\begin{figure*}
\includegraphics[width=\linewidth]{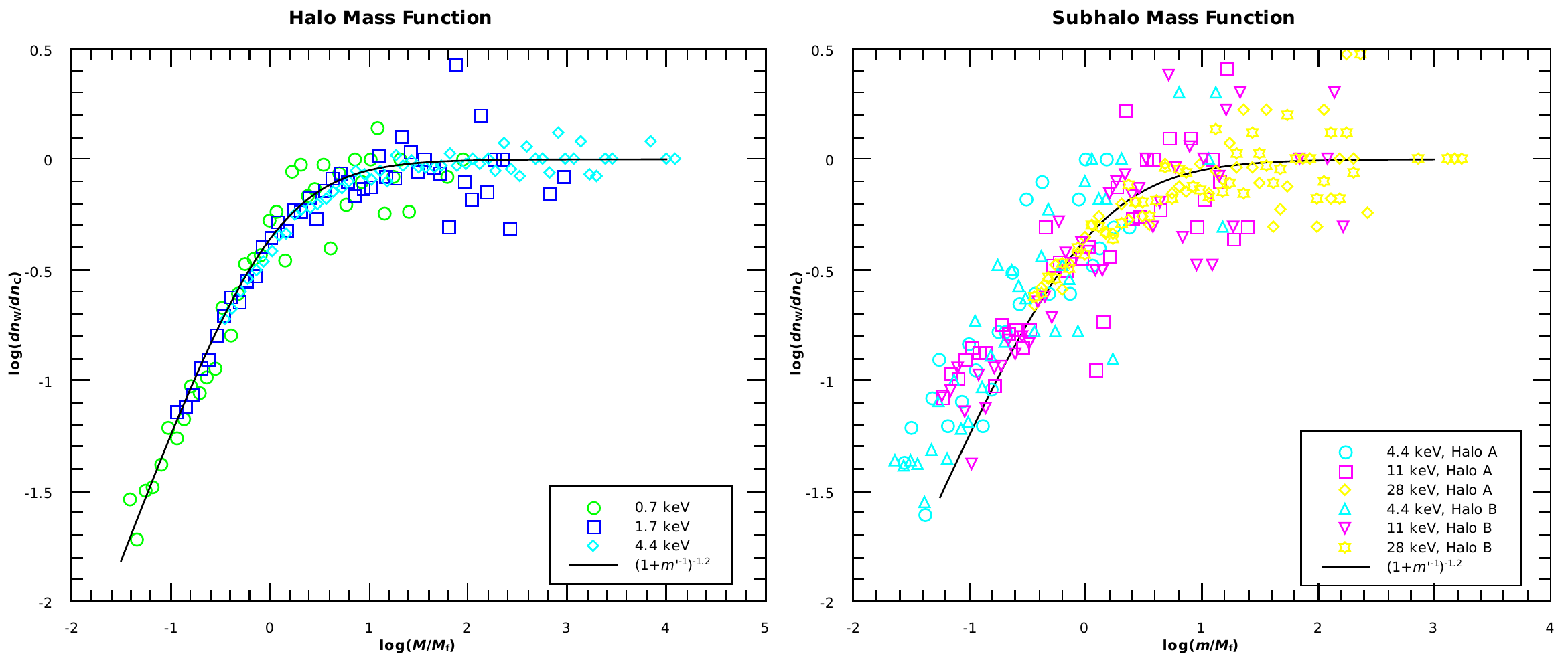}
\caption{\label{fig:WC_mp}The ratio of the WDM to CDM halo and subhalo
  mass functions from our set of simulations is plotted as a function
  of the halo or subhalo mass $m'=m/M_\text{f}$ for several WDM
  particle masses, as given in the legends. The ratio of mass
  functions as a function of $m'$ fits the same form, listed in the
  legend, for both the halo and subhalo mass functions.}
\end{figure*}

For the subhalo mass function, we use our simulations to determine the
change in the subhalo mass function relative to WDM particle mass.  We
want a function such that
\begin{equation}
\frac{dn_\text{W}}{dm} = g\left(m'\right)\frac{dn_\text{C}}{dm},
\end{equation}
where $m'=m/M_\text{f}$ and $M_\text{f}$ is the filtering mass.
To find $g\left(m'\right)$, we take the ratio of the WDM to CDM mass
functions and change variables to $m'$.  The result of this
procedure for both the halo and subhalo mass functions is shown in
Fig.~\ref{fig:WC_mp}.  As expected from \S \ref{haloMF:WDM}, the halo
mass function matches the multiplicative factor given in
Eq.~\eqref{eqn:dn_WDM}.  That is,
\begin{equation} 
\frac{dN_\text{W}/dM}{dN_\text{C}/dM} =
  \left(1+M'^{-1}\right)^{-1.2}, 
\end{equation} 
where $M'=M/M_\text{f}$ of the host halo. When the same procedure is
applied to the subhalo mass function, we find that the data fits the
same expression as for the halo mass function.  As with the host
halos, we use a small scale cut-off for the subhalos.  Our subhalo
mass function changes with WDM particle mass in exactly the same way
as the halo mass function in Eq.~\eqref{eqn:dn_WDM_2}, but using
Eq.~\eqref{eqn:sub_cdm} for the CDM subhalo mass function.  There may
be slight evidence, as seen in Fig.~\ref{fig:WC_mp}, that the mass
function for subhalos is less steep as a function of $m'$.

We normalize the subhalo mass function so that there are the same
number of the largest subhalos (0.01 the halo mass) in CDM and WDM.
The constant $N_0$ from Eq.~\eqref{eqn:sub_cdm} in the WDM model is
related to CDM model by
\begin{equation}
N_{0_\text{W}}=\left(1+100\frac{M_\text{f}}{M}\right)^\eta
  N_{0_\text{C}}.
\end{equation} 
The fraction of the halo mass in subhalos is determined by the
integral in Eq.~\eqref{eqn:frac}.  Note the dependence on the host
halo mass.  The smaller the mass of the host halo, the smaller the
fraction of the mass in subhalos.

\begin{figure}
\includegraphics[width=\linewidth]{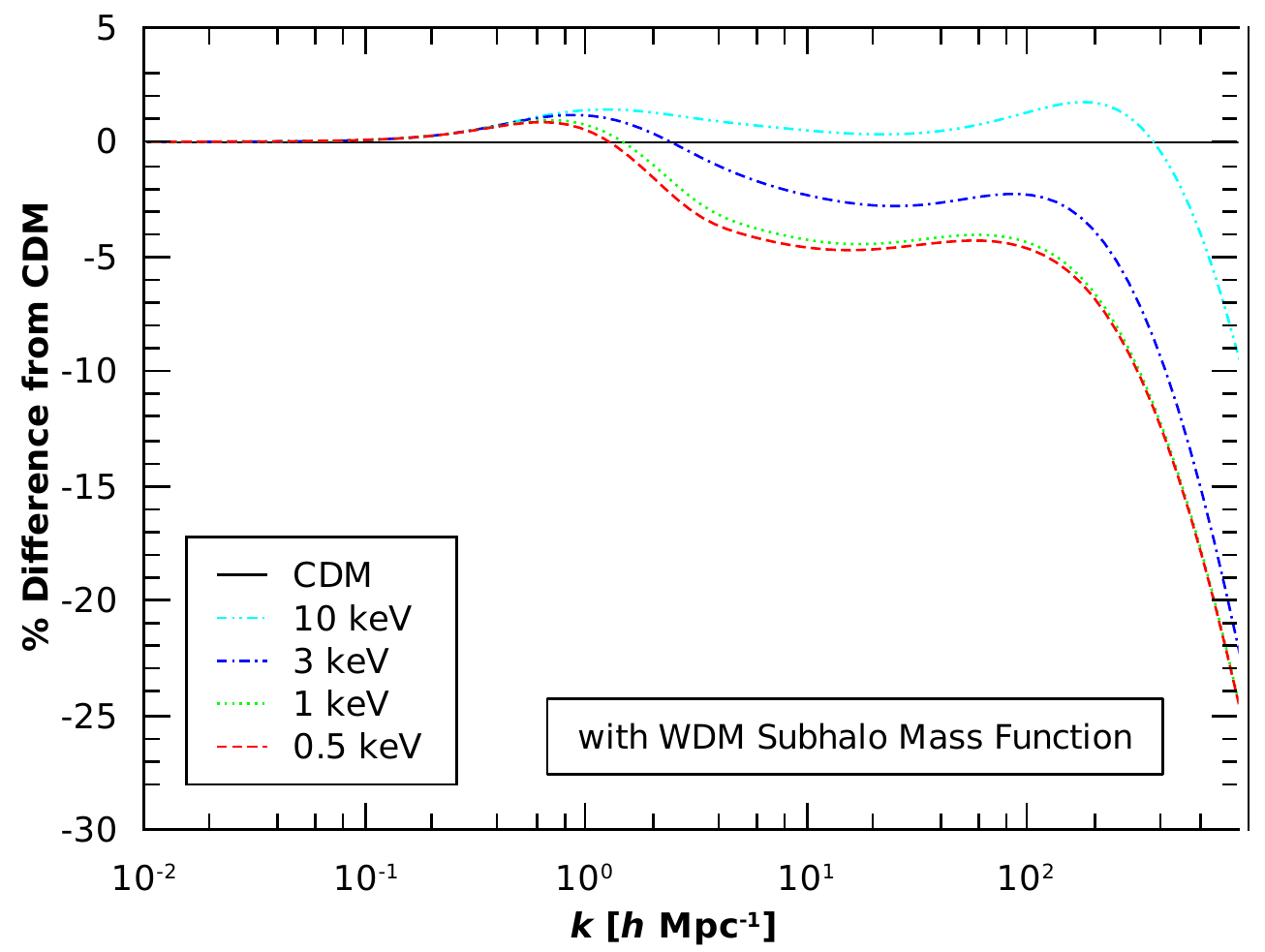}
\caption{\label{img:Pnl_sm}The percent difference between the
  nonlinear matter power spectrum for the CDM model with substructure
  and several WDM particle masses, as listed in the legend, is plotted
  versus wavenumber $k$.  In this graph, the WDM models differ from
  the CDM model in only the subhalo mass function.  The changes to the
  subhalo mass function decrease the fraction of the host halo mass in
  subhalos and increase the fraction of the mass in the smooth
  component of the host halo.  Power is transferred from the smallest
  scales, the subhalos, to somewhat larger scales, the host halo.}
\end{figure}

The effect of the subhalo mass function on the nonlinear matter power
spectrum is shown in Fig.~\ref{img:Pnl_sm}.  As the WDM particle mass
decreases, the slope of the subhalo mass function $\mu$ decreases,
and $M_\text{fs}$ increases.  This decreases the fraction of the
host halo mass in subhalos and increases the fraction of the mass in
the smooth component of the host halo.  This is the reason for the
increase in power in the nonlinear matter power spectrum between 0.1
and 1 $h\ \text{Mpc}^{-1}$.  Power is transferred from the smallest
scales, the subhalos, to somewhat larger scales, the host halo.

\section{\label{smooth}Smooth Background}
\begin{figure}
\includegraphics[width=\linewidth]{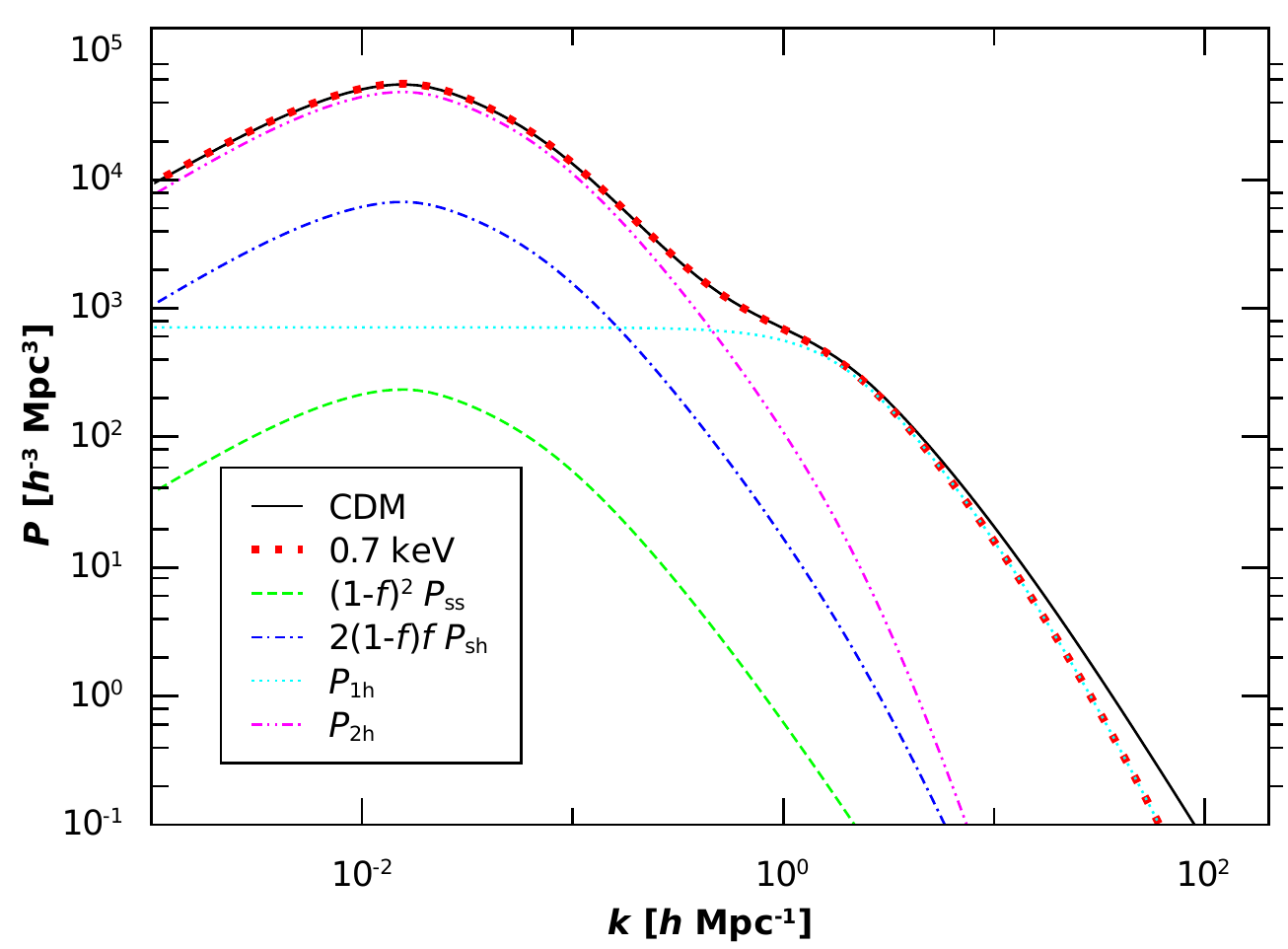}
\caption{\label{img:smooth}The power spectrum for a 0.7 keV particle
  mass, is plotted versus wavenumber $k$.  This graph includes the
  WDM mass function and the effect of a smooth background component to
  the dark matter.  The smooth-smooth, smooth-halo, 1-halo and 2-halo
  power spectrums for a 0.7 keV WDM particle are also shown.  Note
  that the smooth-smooth and smooth-halo terms are much smaller than
  the 2-halo term.}
\end{figure}

The suppression of small mass halos results in some of the dark matter
being in a smooth background component instead of collapsing into
halos.  The fraction of the dark matter in halos is:
\begin{equation}f_\text{h} = \frac{1}{\bar{\rho}_0}\int dM M
  \frac{dn}{dM}.\end{equation} The fraction of the mass in halos is
0.95 for a 10 keV WDM particle, 0.76 for a 3 keV particle, 0.52 for a
1 keV particle and 0.36 for a 0.5 keV particle.  We assume that the
smooth component of the dark matter is related to the linear matter
power spectrum by
\begin{equation} 
b_\text{s}^2 =
\frac{P_\text{ss}\left(k\right)}{P_\text{lin}\left(k\right)},
\end{equation}
where $b_\text{s}$ is derived in Ref.~\cite{Smith:2011ev}:
\begin{equation}
b_\text{s}=\frac{1-f_\text{h}
    b^\text{eff}}{1-f_\text{h}},
\end{equation} and $b^\text{eff}$ is
the effective mass weighted halo bias:
\begin{equation}
b^\text{eff}=\frac{\int dM M \left(dn/dM\right)\
    b\left(M\right)}{\int dM M \left(dn/dM\right)}.
\end{equation}
The power spectrum for two points in the smooth component is
\cite{Smith:2011ev}
\begin{equation}P_\text{ss}\left(k\right)=b_\text{s}^2
  P_\text{lin}\left(k\right),
\end{equation} 
and the correlation between the smooth component and halos is
\begin{equation}
  P_\text{sh}\left(k\right)=\frac{b_\text{s}
    P_\text{lin}\left(k\right)}{\bar{\rho}_\text{h}} 
  \int dM\frac{dn}{dM}M\ u\left(k\mid M\right)\ b\left(M\right),
\end{equation}
where $\bar{\rho}_\text{h}=f_\text{h}\bar{\rho}$.  Then, the full
power spectrum is:
\begin{eqnarray}
  P\left(k\right)&=&\left(1-f_\text{h}\right)^2 P_\text{ss}\left(k\right) \nonumber \\
  &&+2\left(1-f_\text{h}\right)f_\text{h}P_\text{sh}\left(k\right) \nonumber \\
  &&+P_\text{1h}\left(k\right)+P_\text{2h}\left(k\right).
\end{eqnarray}
The full power spectrum and its components for a 0.7 keV WDM particle
are shown in Fig.~\ref{img:smooth}.  Note that the smooth-smooth and
smooth-halo terms are much smaller than the 2-halo term.  Therefore,
these terms should not significantly affect our results.

\begin{figure}
\includegraphics[width=\linewidth]{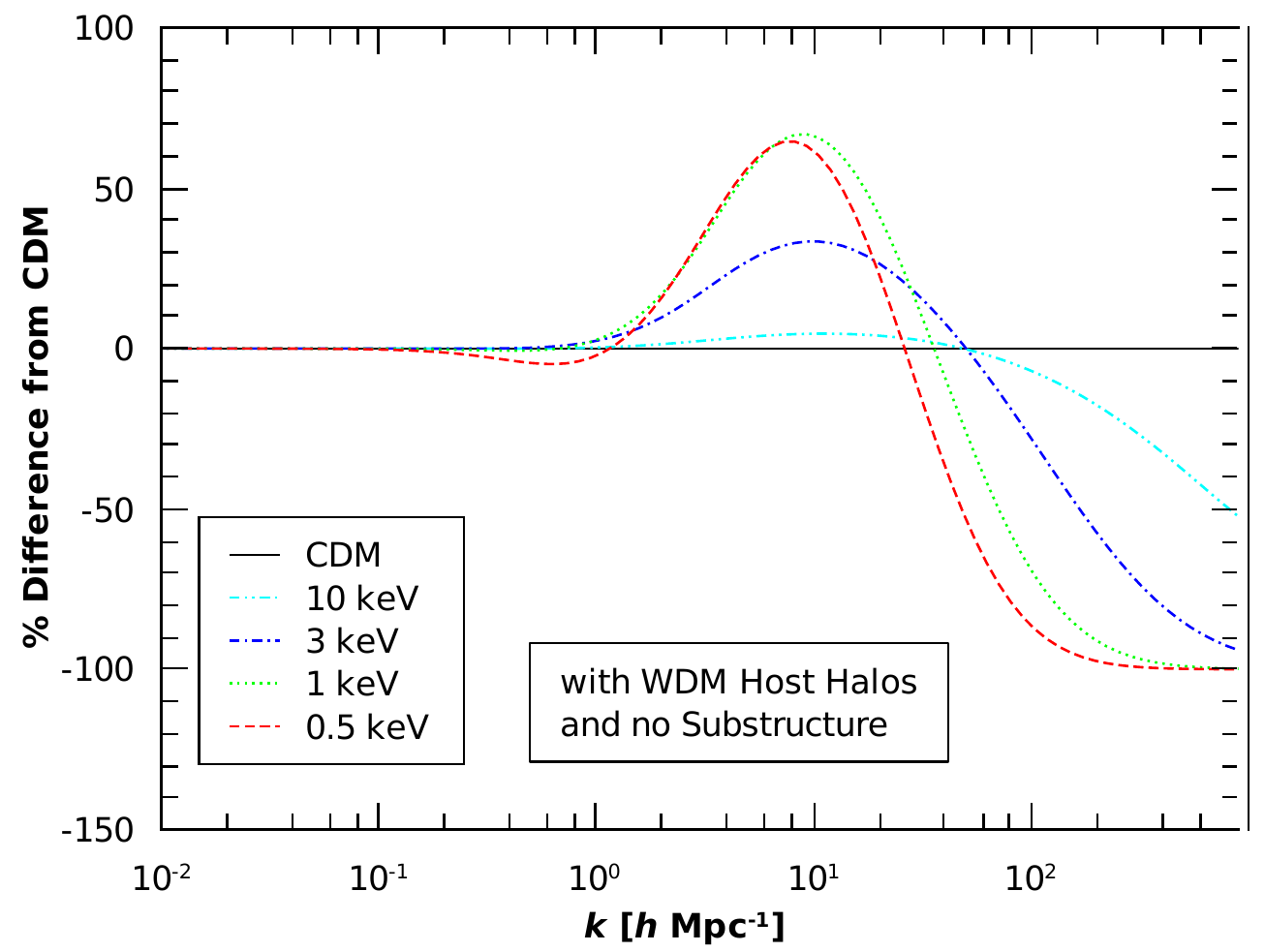}
\caption{\label{img:Pnl_nosub}The percent difference between the
  nonlinear matter power spectrum for the CDM model and several WDM
  particle masses, as listed in the legend, is plotted versus
  wavenumber $k$.  This graph shows the results of changing the
  transfer function, mass function, halo density profile, and
  concentration.  It does not include substructure.  At $k\approx 10
  h$ Mpc$^{-1}$, the strongest effects are from the concentration
  and halo density profile, which increase the power at these scales.
  All four of our alterations to the WDM model in this graph
  contribute to the decrease in power at small scales (large $k$).}
\end{figure}

\begin{figure}
\includegraphics[width=\linewidth]{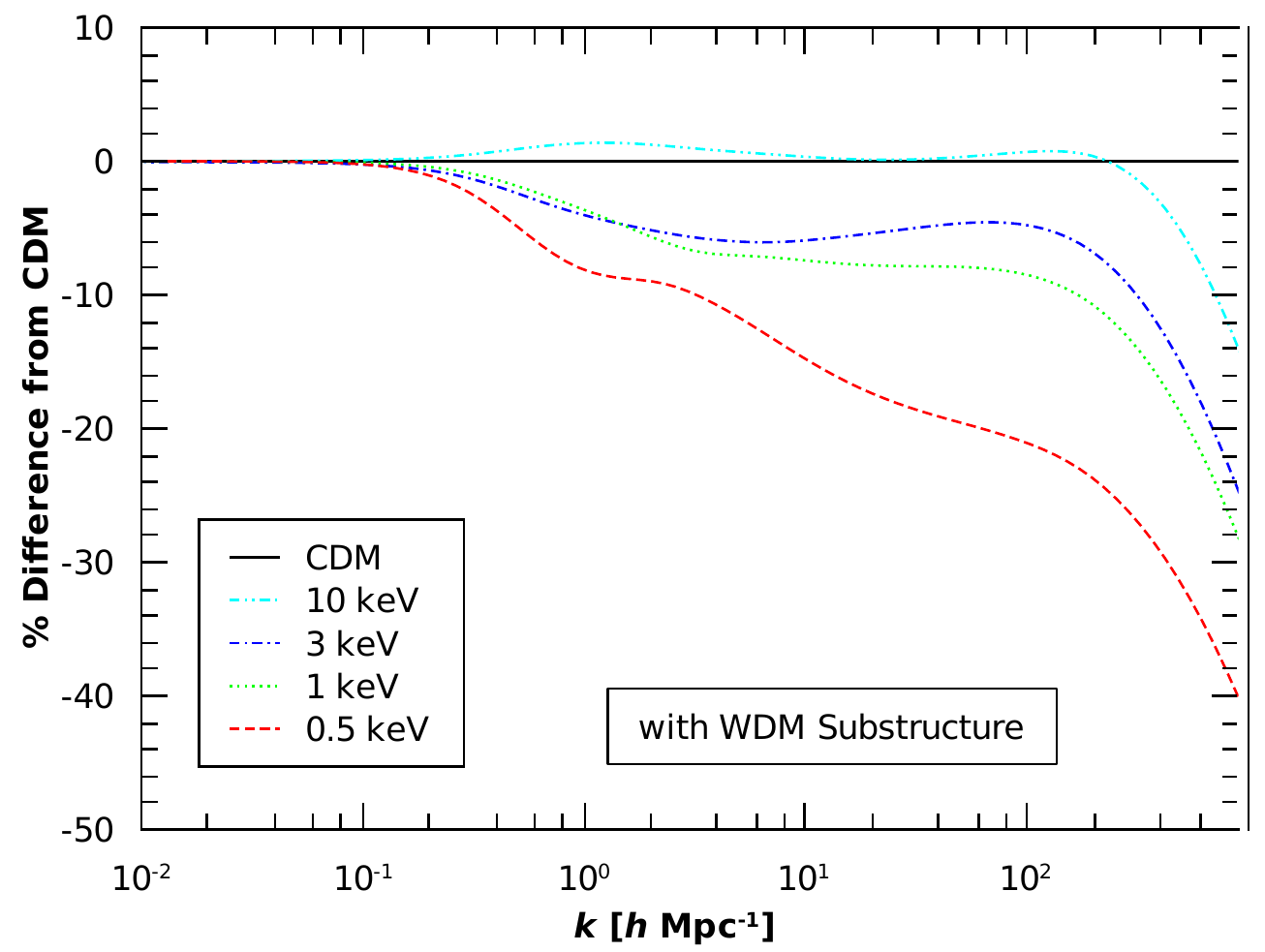}
\caption{\label{img:Pnl_sub}The percent difference between the
  nonlinear matter power spectrum for the CDM model with substructure
  and several WDM particle masses, as listed in the legend, is plotted
  versus wavenumber $k$.  This graph shows the results of changing
  the transfer function, the subhalo mass function, subhalo density
  profile, and subhalo concentration.  The dominant effect here,
  especially for the 10 keV WDM particle, is from the subhalo density
  profile.  For the 0.5 keV WDM particle, the largest effect is the
  decrease in power from the changes to the subhalo concentration.}
\end{figure}

\begin{figure}
\includegraphics[width=\linewidth]{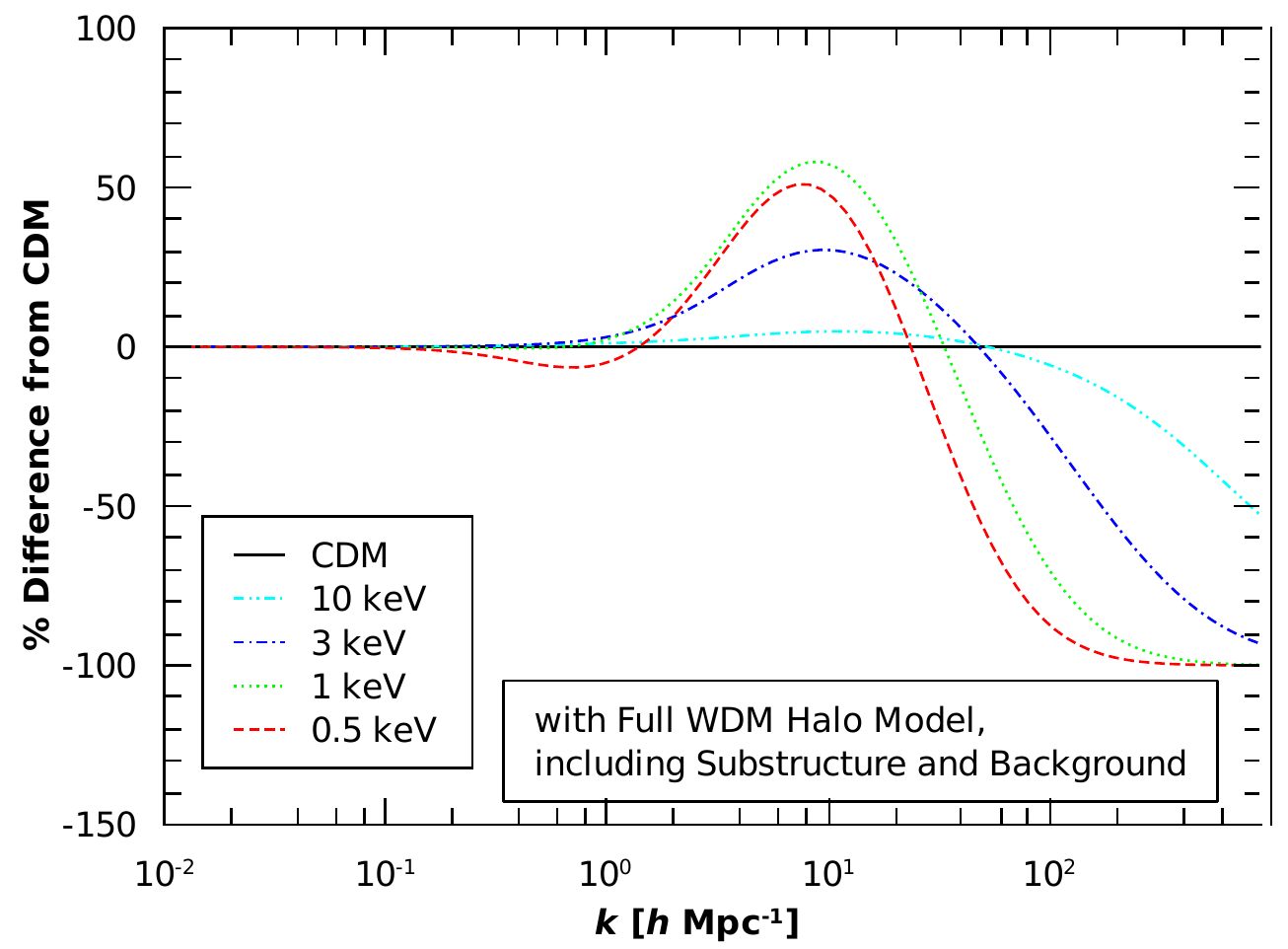}
\caption{\label{img:Pnl_all}The percent difference between the
  nonlinear matter power spectrum for the CDM model with substructure
  and several WDM particle masses, as listed in the legend, is plotted
  versus wavenumber $k$.  This graph includes {\it all} of the
  adjustments to the WDM model detailed in this paper.  This figure
  greatly resembles Fig.~\ref{img:Pnl_nosub}, with the increase in
  power at $k\approx 10 h$ Mpc$^{-1}$.  Therefore, the largest
  effects are from the changes to the host halo properties, with
  changes to the substructure and the addition of a smooth background
  component to the dark matter being sub-dominant.}
\end{figure}

\section{\label{res}Results}

Fig.~\ref{img:Pnl_nosub} is a graph of the percent difference of the
nonlinear matter power spectrum in our WDM model with all effects
except substructure.  That is, the transfer function, mass function,
halo density profile, and concentration are altered as described in
the previous sections.  At $k\approx 10 h$ Mpc$^{-1}$, the
strongest effects are from the concentration and halo density profile,
which increase the power at these scales.  All four of our alterations
to the WDM model in this graph contribute to the decrease in power at
small scales (large $k$).

Fig.~\ref{img:Pnl_sub} is a graph of the percent difference of the
nonlinear matter power spectrum in our WDM model, including only the
effects of WDM on the substructure and the transfer function, compared
to the CDM model with substructure.  The dominant effect at the 10
$h\ \text{Mpc}^{-1}$ scale, especially for the 10 keV WDM particle,
is from the subhalo mass function.  For the 0.5 keV WDM particle, the
largest effect is the decrease in power from the changes to the
subhalo concentration.

Fig.~\ref{img:Pnl_all} is a graph of the percent difference of the
nonlinear matter power spectrum for our WDM model with {\it all} of
the effects discussed in this paper compared to the CDM model with
substructure.  The features in the changes of the nonlinear matter
power spectrum greatly resemble those in Fig.~\ref{img:Pnl_nosub}.
Therefore, the largest effects are from the changes to the main halo
in WDM, with changes to the substructure and the addition of a smooth
background component to the dark matter being smaller effects.

\section{Conclusion}
\label{concl}
We have presented our work on a broad assessment of the
effects of WDM in the halo model of large-scale structure in two-point
statistics, employing results from our own set of simulations as well
as previously published work.  This framework may be incorporated in a
number of applications, including forecasts for large scale structure
measures such as weak lensing~\cite{Markovic:2010te,Smith:2011ev}, in
galaxy clustering two-point function measures of the power
spectrum~\cite{vandenBosch:2003nk,*Abazajian:2004tn}.   We have
included the effects to the linear matter power spectrum, halo density
profile, halo concentration relation, halo mass function, subhalo
density profile, subhalo mass function and biasing of the smooth dark
matter component in the case of WDM.

We find a drastic difference in the nonlinear matter power spectrum
predicted by the halo model between WDM and CDM even for reasonably
``cold'' WDM particle masses ($m_s = 10\rm\ keV$) at the smallest
scales, as expected.  We show that, counter-intuitively, WDM produces
an enhancement of power at intermediate scales due to the softening of
halo density concentrations and profiles.  Since cored profiles are
not found in WDM simulations, nor expected in analytic Gaussian peak
statistics of WDM~\cite{deNaray:2009xj,VillaescusaNavarro:2010qy}, we
do not include cored profiles, which differs our results from
Ref.~\cite{Smith:2011ev}. Also different from that work, we also
include the effects of halo substructure, subhalo density profiles and
subhalo mass functions.  Ref.~\cite{Markovic:2010te} only included the
effects of the halo profile and halo mass function, and assumed all
mass to be in halos, which is not the case in WDM.

We find that the host halo effects dominate the overall effects of WDM
versus CDM in the nonlinear regime.  Though our detailed results
differ from previous work, the overall magnitude of the effects in the
intermediate scale ($k\sim 10 - 100\ h \rm\ Mpc^{-1}$) regimes relevant
for weak lensing are comparable, and therefore, the estimates for weak
lensing sensitivity in that work are likely not drastically changed
when incorporating our model in a weak lensing forecast.

The halo model has broad applications to observable cosmological
statistics.  With the help of the nonlinear structure framework of WDM
versus CDM like that presented here, the observed large to small-scale
clustering of matter, gas, and galaxies may be able to shed light on
the nature of dark matter and its primordial origin.

\begin{acknowledgments}
  We would like to thank Manoj Kaplinghat for detailed comments on the
  manuscript and An\v{z}e Slosar for useful discussions.  RMD and KNA
  are supported by NSF Grant 07-57966 and NSF CAREER Grant
  09-55415. EP acknowledges support under the Edison Memorial Graduate
  Training Program at the Naval Research Laboratory.
\end{acknowledgments}

\bibliography{master}

\end{document}